%% file: main.tex
\documentclass[11pt,preprint]{elsarticle}
\usepackage{adjustbox}
\usepackage{amsthm}
\usepackage{amsmath}
\usepackage{amsfonts,amssymb,graphicx,tikz,bm,blindtext}
\usepackage{pifont}
\usepackage{booktabs}
\usepackage{float}
\usepackage[margin=1in]{geometry}
\usepackage[utf8]{inputenc}
\usepackage{mathtools}
\usepackage[separate-uncertainty=true]{siunitx}
\usepackage{subfigure}
\usepackage{xcolor}
\usepackage[colorlinks=true]{hyperref} 
\usepackage{arydshln}
\usepackage{cleveref}
\usepackage{comment}
\usepackage{lineno}

\graphicspath{{figs/}}
\journal{Journal of \LaTeX\ Templates}

\newtheorem{thm}{Theorem}[section]

\numberwithin{equation}{section}
\newcommand{\mean}[1]{\mathbb{E}\lbrack #1\rbrack}
\newcommand{\var}[1]{ {\rm Var} \lbrack #1\rbrack}

\newcommand{\norm}[1]{\left\Vert#1\right\Vert}  

\newcommand{\set}[1]{\left\{#1\right\}}
\newcommand{\Real}{\mathbb R}

\newcommand{\cF}{\mathcal{F}}
\newcommand{\cG}{\mathcal{G}}

\newcommand{\cI}{\mathcal{I}}

\newcommand{\cL}{\mathcal{L}}

\newcommand{\cU}{\mathcal{U}}
\newcommand{\cV}{\mathcal{V}}

\crefformat{section}{\S#2#1#3} 
\crefformat{subsection}{\S#2#1#3}
\crefformat{subsubsection}{\S#2#1#3}

\begin{document}


\begin{frontmatter}

\title{A comprehensive and fair comparison of two neural operators \\
(with practical extensions) based on FAIR data}

\author[1]{Lu Lu\fnref{fn1}}
\author[2]{Xuhui Meng\fnref{fn1}}
\author[2]{Shengze Cai\fnref{fn1}}
\author[3]{Zhiping Mao}
\author[2]{Somdatta Goswami}
\author[4]{Zhongqiang Zhang}
\author[2,5]{George Em Karniadakis\corref{mycorrespondingauthor}}
\cortext[mycorrespondingauthor]{Corresponding author. Email: george\_karniadakis@brown.edu}
\fntext[fn1]{These authors contributed equally to this work.}

\address[1]{Department of Chemical and Biomolecular Engineering, University of Pennsylvania}
\address[2]{Division of Applied Mathematics, Brown University}
\address[3]{School of Mathematical Sciences, Xiamen University}
\address[4]{Department of Mathematical Sciences, Worcester Polytechnic Institute}
\address[5]{School of Engineering, Brown University}

\begin{abstract}
Neural operators can learn nonlinear mappings between function spaces and offer a new simulation paradigm for real-time prediction of complex dynamics  
for realistic diverse applications as well as for system identification in science and engineering. 
Herein, we investigate the performance of two neural operators, which have shown promising results so
far, and we develop new practical extensions that will make them more accurate and robust and importantly more suitable for industrial-complexity
applications. The first neural operator, {\em DeepONet}, was published in 2019 \cite{lu2019deeponet}, and its original architecture was based on the
universal approximation theorem of Chen \& Chen \cite{chen1995universal}. The second one, named Fourier Neural Operator or {\em FNO}, was published in 2020 \cite{li2020fourier}, and it is based on parameterizing the integral kernel in the Fourier space.
DeepONet is represented by a summation of products of neural networks (NNs), corresponding to the branch NN for the input function and the trunk NN for the output function; both NNs are general architectures, e.g., the branch NN can be replaced with a CNN or a ResNet. According to \cite{kovachki2021universal}, FNO in its continuous form can be viewed as a DeepONet with a specific architecture of the branch NN and a trunk NN represented by a trigonometric basis.
In order to compare FNO with DeepONet for realistic setups,  we develop several extensions of FNO that can deal with complex geometric domains
as well as mappings where the input and output function spaces are of different dimensions. 
We also endow DeepONet with special features that provide inductive bias and 
accelerate training, and we present a faster implementation of DeepONet with cost comparable to the computational cost of FNO, 
which is based on the Fast Fourier Transform.

We consider 16 different benchmarks to demonstrate the relative performance of the two neural operators, including instability wave analysis 
in hypersonic boundary layers, prediction of the vorticity field of a flapping airfoil, porous media simulations in complex-geometry domains, etc. 
The performance of DeepONet and FNO is comparable for relatively simple settings, but for complex geometries and especially noisy data, the performance of FNO deteriorates greatly. For example, for the instability wave analysis with only $0.1\%$ noise added to the input data, the error of FNO increases 10,000 times making it inappropriate for such important applications, while there is hardly any effect of such noise on the DeepONet. This behavior of FNO may be related to an inherent mapping instability between function spaces, and hence future work should address this important issue. We also compare theoretically the two neural operators and obtain similar error estimates for DeepONet and FNO under the same regularity assumptions.
\end{abstract}


\end{frontmatter}


\section{Introduction}

While there have been rapid developments in machine learning in the last 20 years and there is currently plenty of euphoria and admiration about the so-called ``unreasonable effectiveness of deep learning in artificial intelligence''
\cite{sejnowski2020unreasonable},
the development of physics-informed machine learning and its application to scientific applications is relatively recent \cite{karniadakis2021physics}. 
Physics-informed neural networks (PINNs) were introduced first in 2017 \cite{raissi2017physics,raissi1711physics} for forward, inverse and hybrid problems, and since then there have also been rapid developments in this area \cite{lu2021deepxde,pang2019fpinns,zhang2019quantifying,zhang2020learning,lu2021physics,meng2020ppinn,jagtap2020extended,yu2021gradient,cai2021physics}, although the achievements so far in scientific and engineering applications have been modest compared to applications in imaging, speech recognition and natural language processing. According to some estimates, last year alone close to 100,000 papers on machine learning were published, all of which were based on the assumption of the universal function approximation of 
neural networks --- a theoretical work that goes back to the early 1990s~\cite{cybenko1989approximation,hornik1989multilayer}.

At about the same time, a much less known theorem of Chen \& Chen \cite{chen1995universal} on the universal operator approximation by single-layer neural networks was developed but remained largely unknown until our work on DeepONet in 2019 \cite{lu2019deeponet} with subsequent theoretical and computational extensions in \cite{lu2021learning}. Unlike function regression, operator regression aims to map infinite-dimensional functions (inputs) to infinite-dimensional functions (outputs). The work in
\cite{lu2021learning} extended the theorem of Chen \& Chen \cite{chen1995universal} to deep neural networks, which are more expressive and break the curse of dimensionality in the input space.
More importantly, from the computational point of view, the new paradigm of operator regression allows for simulating the dynamics of complex nonlinear systems, e.g., fluid flows, corresponding to different boundary and initial conditions without the need for retraining the neural network. As noted independently by DeepMind researcher Irina Higgins \cite{higgins2021generalizing},  
``Once DeepONet is trained, it can be applied to new input functions, thus producing new results substantially faster than numerical solvers. Another benefit of DeepONet is that it can be applied to simulation data, experimental data or both, and the experimental data may span multiple orders of magnitude in spatio-temporal scales, thus allowing scientists to estimate dynamics better by pooling the existing data.'' DeepONet can be applied to partial differential equations (PDEs) but also to learning explicit mathematical operators, e.g., integration, fractional derivatives, Laplace transforms, etc. \cite{lu2021learning}.

Another parallel effort on operator regression started in 2020 with a paper on a graph kernel network (GKN) for PDEs \cite{li2020neural}. The authors represented the infinite-dimensional mapping by composing nonlinear activation functions and a class of integral operators with the kernel integration computed by message passing on graph networks. Unfortunately, GKN was of limited use as it was shown to be unstable with the increase of the number of hidden layers \cite{yu2021nonlocal}.
A different architecture was then proposed by the same group in \cite{li2020fourier}, where they formulated the operator regression by parameterizing the integral kernel directly in Fourier space. They demonstrated very good accuracy and efficiency for relatively simple settings, including the Burgers' equation, Darcy flow, and the Navier-Stokes equations. However, they made the claim that ``The Fourier neural operator is the first ML-based method to successfully model turbulent flows with zero-shot super-resolution. It is up to three orders of magnitude faster compared to traditional PDE solvers. Additionally, it achieves superior accuracy compared to previous learning-based solvers under fixed resolution.'' This is of course erroneous as the flow considered was a 
simple smooth laminar flow, and their comparison with DeepONet was not properly conducted. In fact, as we will demonstrate herein, DeepONet can achieve similar and
in fact better accuracy for the same or similar benchmarks presented in \cite{li2020fourier} and even superior accuracy in realistic problems involving complex-geometry domains and noisy input data.

In this paper, we present a very systematic and transparent study of comparing the performance of DeepONet and FNO for 16 different benchmarks 
selected carefully to highlight both the advantages and the limitations of the two neural operators. According to an independent work by
\cite{kovachki2021universal}, FNO  in its continuous form can be viewed as a DeepONet with a specific architecture of the branch and a trunk represented by a trigonometric basis.
Hence, in the current work we introduce two significant enhancements to FNO so that they can deal with mappings of different dimensionality as well as with complex-geometry domains, so that we can make sensible comparisons with DeepONet, which is a general neural operator. We also introduce 
various enhancements to DeepONet to accelerate its training and increase its accuracy, introducing for example the POD modes in the trunk net, obtained readily  from the available training datasets. In addition to computational tests, we also perform a theoretical comparison of DeepONet versus FNO, following the 
published work on the theory of DeepONet in \cite{chen1995universal,lanthaler2021error,deng2021convergence,yu2021arbitrary}, and on the more recent theory of FNO
in \cite{kovachki2021universal,kovachki2021neural}. On this point, it is worth noted that DeepONet was based from the onset on the theorem of Chen \& Chen \cite{chen1995universal}, whereas the formulation of FNO was not theoretically justified originally, and the recent theoretical work covers only invariant kernels. There are also other methods for operator regression such as \cite{bhattacharya2020model,trask2019gmls,you2021data,patel2021physics}, but in this work we only consider DeepONet and FNO.

In the following, we summarize the new contributions of the current work in addition to designing 16 different benchmarks and
obtaining new results for both DeepONet and FNO along with their new extensions.
More specifically, the new developments for DeepONet are the following:
\begin{itemize}
    \item We introduce extra features in the trunk net and the branch net.
    \item We impose hard-constraints for Dirichlet and periodic boundary conditions via a modified trunk net.
    \item We develop a new extension, POD-DeepONet, that employs the POD modes of the training data as the trunk net.
    \item We analyze and test a new DeepONet scaling that leads to accuracy improvement.
    \item We extend DeepONet to deal with multiple outputs.
    \item We present a new fast implementation of DeepONet, comparable to FNO for similar settings.
\end{itemize}
Similarly, the new developments for FNO are the following:
\begin{itemize}
\item dFNO+: We extend FNO to nonlinear mappings with inputs and outputs defined on different domains.
\item gFNO+: We extend FNO to nonlinear mappings with inputs and outputs defined on a complex geometry.
\item We add extra features by using them as extra network inputs. 
\end{itemize}
Moreover, we employ normalization of both inputs and outputs for DeepONet and FNO and demonstrate its effect.

In our comparative studies, we designed benchmarks with important features typically encountered in real world applications, such as
complex-geometry domains, non-smooth solutions, unsteadiness, and noisy data. On the theoretical side, our contribution is on 
developing error estimation of the network size by emulating Fourier methods both for DeepONet and FNO. We aim to make this 
study, including codes and data, accessible to all, and to the degree possible we have followed the FAIR (Findability, Accessibility, Interoperability, and Reusability) guiding principles for 
scientific data management and stewardship \cite{wilkinson2016fair}.

The paper is organized as follows. In the next section, we define the problem setup and the data types. In Section~\ref{sec:network}, we 
describe the DeepONet and FNO architectures as well as their extensions. In Section~\ref{sec:theory}, we present a theoretical comparison of the 
approximation theorems for the two neural operators and in addition we compare their error estimates for the solution of the Burgers' equation.
In Section~\ref{sec:results}, we compare the performance of DeepONet and FNO for 16 different benchmarks listed in Table~\ref{tab:problems}. Finally, we conclude with a summary in Section~\ref{sec:summary}. The Appendix includes the proof of the theorem presented in the main text, description of the data sets, description of the architectures employed, and data generation for the Darcy problems and the cavity flows. 

\section{Operator learning}

We first present the problem setup of operator learning and then discuss some important aspects of the dataset used for learning.

\subsection{Problem setup}

We consider a physical system, which involves multiple functions. Among these functions, we are usually interested in one function and aim to predict this function from other functions. For example, when the physical system is described by partial differential equations (PDEs), these functions typically include the PDE solution $u(x, t)$, the forcing term $f(x, t)$, the initial condition (IC) $u_0(x)$, and the boundary conditions (BCs) $u_b(x, t)$, where $x$ and $t$ are the space and time coordinates, respectively. 

We denote the input function by $v$ defined on the domain $D \subset \mathbb{R}^d$, e.g., $f(x, t)$ or $u_0(x)$,
$$v: D \ni x \mapsto v(x) \in \mathbb{R},$$
and denote the output function by $u$ defined on the domain $D' \subset \mathbb{R}^{d'}$
$$u: D' \ni \xi \mapsto u(\xi) \in \mathbb{R}.$$
Let $\mathcal{V}$ and $\mathcal{U}$ be the spaces of $v$ and $u$, respectively, and $D$ and $D'$ could be two different domains. Then, the mapping from the input function $v$ to the output function $u$ is denoted by an operator
$$\mathcal{G}: \mathcal{V} \ni v \mapsto u \in \mathcal{U}.$$
In this study, we aim to approximate $\mathcal{G}$ by neural networks and train the network from a training dataset $\mathcal{T} = \left\{\left(v^{(1)}, u^{(1)}\right), \left(v^{(2)}, u^{(2)}\right), \cdots, \left(v^{(m)}, u^{(m)}\right)\right\}$.

\subsection{Data types}

Deep learning is a powerful method for leveraging big data, but in many science and engineering problems, the available data is not ``big'' enough to ensure accuracy and reliability of deep learning models. What we may have instead is ``dinky, dirty, dynamics, and deceptive data'' as fist characterized by Alexander Kott, chief of the Network Science Division of the US Army Research Laboratory \cite{maupin2019us}. On the other hand, scientific data should meet principles of \textbf{findability, accessibility, interoperability, and reusability} (FAIR) \cite{wilkinson2016fair}.

In addition, we have to facilitate multi-modality input data that may come from diverse sources, e.g., static images, videos, Schlieren photography, particle image velocimetry (PIV), particle tracking velocimetry (PTV),  radar and satellite images, MRI, CT, X-ray, two-photon microscopy, satellite, synthetic/simulated data, scattered unstructured opportunistic data, etc. We also have to facilitate multi-fidelity data that not only include multi-resolution data but also data corresponding to different levels of physical complexity, and hence training of NN requires special multi-fidelity methods such as in \cite{meng2020composite,meng2021multi,lu2020extraction}.
Noisy data are omnipresent and effective NN training and inference should be stable to noise and provide answers with uncertainty quantification using, e.g., Bayesian NN as in \cite{yang2021b,meng2021multi,olivier2021bayesian,meng2021learning}.

All the codes and data will be available on Github \url{https://github.com} upon publication of our paper.

\section{Operator regression networks}
\label{sec:network}

We first introduce the two neural operators, namely the deep operator network (DeepONet)~\cite{lu2019deeponet,lu2021learning} and the Fourier neural operator (FNO)~\cite{li2020fourier}, and subsequently we propose several extensions of these two methods.

\subsection{DeepONet}

Next, we provide an introduction to the vanilla\footnote{In this work, ``vanilla'' means the standard or unmodified version.} DeepONet. Although vanilla DeepONet has demonstrated good performance in diverse applications \cite{lu2021learning,lin2021operator,cai2021deepm,di2021deeponet,mao2021deepm,lin2021seamless,goswami2021physics,wang2021learning,yin2021simulating}, here, we propose several extensions of DeepONet to achieve better accuracy and faster training.

\subsubsection{Vanilla DeepONet}

\begin{figure}[htbp]
    \centering
    \includegraphics[width=\textwidth]{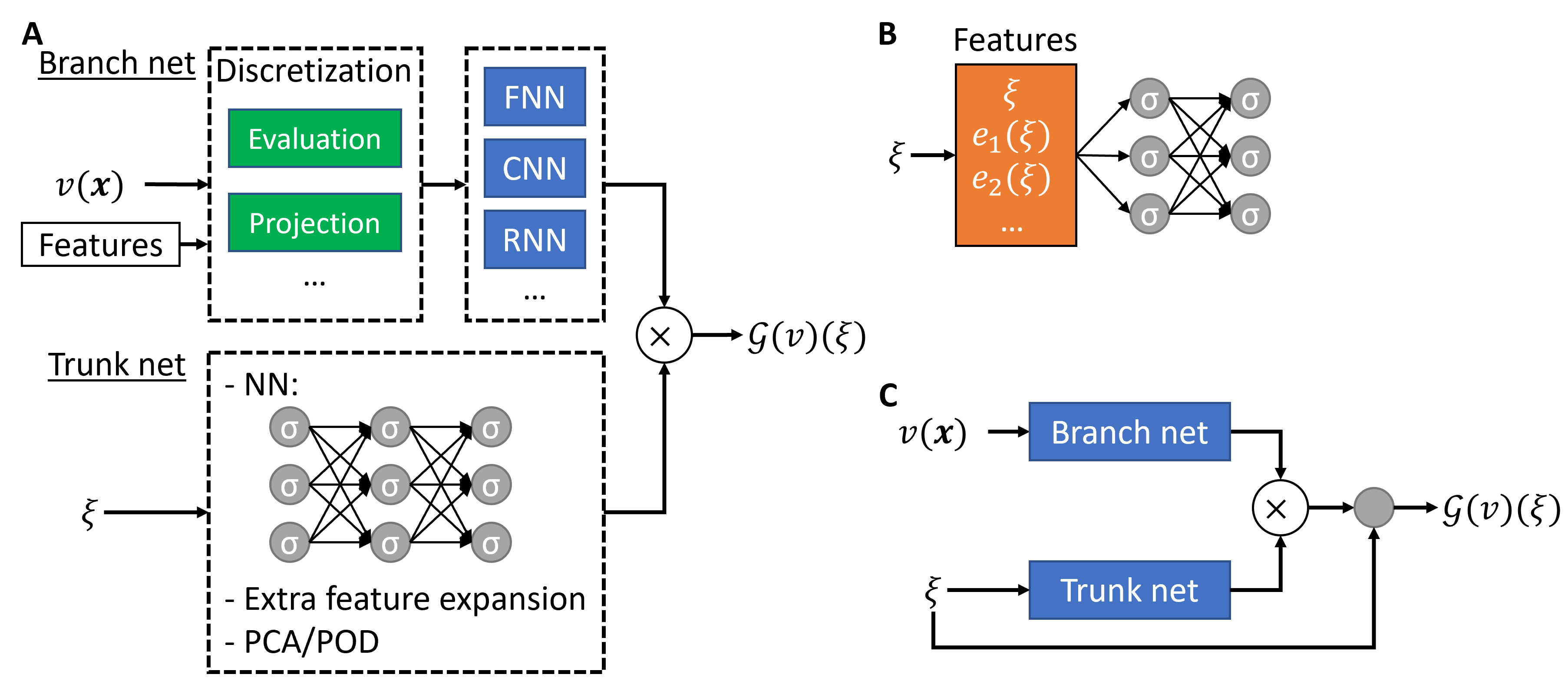}
    \caption{\textbf{Architecture of DeepONet.} (\textbf{A}) DeepONet architecture. If the trunk net is a feed-forward neural network, then it is a vanilla DeepONet. (\textbf{B}) Feature expansion of the trunk-net input. Periodic BCs can also be strictly imposed into the DeepONet by using the Fourier feature expansion. (\textbf{C}) Dirichlet BCs are strictly enforced in DeepONet by modifying the network output.}
    \label{fig:deeponet}
\end{figure}

Four slightly different versions of DeepONet have been developed in Ref.~\cite{lu2021learning}, but in this study we use the stacked DeepONet with bias, which exhibits the best performance in practice among all our versions of DeepONet. First of all, to work with the input function numerically, we need to discretize the input function $v$ and represent $v$ in a finite-dimensional space (Fig.~\ref{fig:deeponet}A). We could evaluate $v$ at a set of locations $\{x_1, x_2, \dots, x_m\}$, i.e., the pointwise evaluations of $[v(x_1), v(x_2), \dots, v(x_m)]$. Many alternative representations could be applied. For example, we could project $v$ to a set of basis functions and an approximation of $v$ is then characterized/parameterized by the coefficients.

In this work, we choose the evaluation approach to dicretize $v$. One advantage of DeepONet is that we do not dicretize the output function $u$ because of the following network design. A DeepONet has two sub-networks: a ``trunk'' network and a ``branch'' network (Fig.~\ref{fig:deeponet}A). The trunk net takes the coordinates $\xi \in D'$ as the input, while the branch net takes the discretized function $v$ as input. Finally, the output of the network is expressed as:
$$\mathcal{G}(v)(\xi) = \sum_{k=1}^p b_{k}(v) t_{k}(\xi) + b_0,$$
where $b_0 \in \mathbb{R}$ is a bias. $\{b_1, b_2, \dots, b_p\}$ are the $p$ outputs of the branch net, and $\{t_1, t_2, \dots, t_p\}$ are the $p$ outputs of the trunk net.

We note that DeepONet is a high-level framework without restricting the branch net and the trunk net to any specific architecture. As $\xi$ is ususally low dimensional, a standard FNN is commonly used as the trunk net. The choice of the branch net depends on the structure of the input function $v$, and it can be chosen as a FNN, ResNet, CNN, RNN, or a graph neural network (GNN), etc. For example, if the discretization of $v$ is on an equispaced 2D grid, then a CNN can be used; if the discretization of $v$ is on a unstructured mesh, then a GNN can be used.

\subsubsection{Feature expansion}
\label{sec:feature}

In many applications, we not only have a dataset, but also have some prior knowledge about the underlying system. For example, we may know that the output functions have an oscillating nature or the functions have a fast decay. It is then beneficial to encode the knowledge directly into DeepONet by modifying the DeepONet architecture. The specific way to encode the knowledge is problem dependent, and here we introduce two ways of feature expansion either in the trunk net or in the branch net.

\paragraph{Feature expansion in the trunk net} If we know some features of the output function $u(\xi)$, then we can construct a feature expansion ($e_1(\xi), e_2(\xi), \dots$) for the trunk net input (Fig.~\ref{fig:deeponet}B), which was originally proposed and used in PINNs \cite{yazdani2020systems,lu2021physics}. For example, as proposed in \cite{di2021deeponet}, for the problem with oscillating solutions, we can first perform a harmonic feature expansion on the input $\xi$ of the trunk network as
$$\xi \mapsto (\xi, \cos(\xi), \sin(\xi), \cos(2\xi), \sin(2\xi), \dots),$$
which then becomes input to the trunk net. In this example, the features have analytical formulas. In general, we may not have the analytical formulas of features, and we can also use numerical values of the features as the extra input of the trunk net. For example, in dynamical systems, we can use some historical data as the feature. We note that here we usually still keep $\xi$ as a feature.

\paragraph{Feature expansion in the branch net} If the feature is a function of $x$, then we cannot encode it in the trunk net, and instead we can use the feature as an extra input function of the branch net (Fig.~\ref{fig:deeponet}A).

\subsubsection{Hard-constraint boundary conditions}
\label{sec:deeponet-bc}

In some applications, we may know the boundary conditions of the system. Here, we introduce how to enforce the Dirichlet BCs and Periodic BCs as hard constraints in DeepONet.

\paragraph{Dirichlet BCs.} Enforcing Dirichlet BCs in neural networks has been widely used in PINNs \cite{lu2021physics}. Let us consider a Dirichlet BC for the DeepONet output:
$$\mathcal{G}(v)(\xi) = g(\xi), \quad \xi \in \Gamma_D,$$
where $\Gamma_D$ is a part of the boundary. To make the DeepONet output satisfy this BC automatically, we construct the solution as
$$\mathcal{G}(v)(\xi) = g(\xi) + \ell(\xi)\mathcal{N}(v)(\xi),$$
where $\mathcal{N}(v)(\xi)$ is the product of the branch and trunk nets, and $\ell(\xi)$ is a function satisfying the following condition:
\begin{equation*}
    \left\{ \begin{array}{l}
    \ell(\xi) = 0, \quad \xi \in \Gamma_D, \\
    \ell(\xi) > 0, \quad \text{otherwise}.
    \end{array}
    \right.
\end{equation*}
Here, we assume $g(\xi)$ is well defined for any $\xi$, otherwise we can construct a continuous extension for $g$.

\paragraph{Periodic BCs} Here, we first introduce how to enforce periodic BCs in neural networks in 1D~\cite{lu2021physics,dong2021method} by constructing special feature expansion as discussed in Section~\ref{sec:feature}, and then discuss how to extend it to 2D. If the solution $u(\xi)$ is periodic with respect to $\xi$ of the period $P$, then $u(\xi)$ can be represented well by the Fourier series. Hence, we can replace the network input $\xi$ with Fourier basis functions, i.e., the features in Fig.~\ref{fig:deeponet}B are 
$$\left\{1, \cos\left(\omega \xi\right), \sin\left(\omega \xi\right), \cos\left(2\omega \xi\right), \sin\left(2\omega \xi\right), \cdots\right\}$$
with $\omega = \frac{2\pi}{P}$. Compared to the aforementioned feature expansion, here we do not have the feature $\xi$ any more. Because each Fourier basis function is periodic, it is easy to prove that the DeepONet output $\mathcal{G}(v)(\xi)$ is also periodic \cite{dong2021method}. The number of Fourier features to be used is problem dependent, and we may use as few as the first two Fourier basis function $\{\cos\left(\omega \xi\right), \sin\left(\omega \xi\right)\}$.

Next, we discuss the 2D case, and with a slight abuse of the notation for $x$ we denote $\xi = (x, y) \in \mathbb{R}^2$. The basis functions of the Fourier series on a 2D square are: 
$$\left\{ \cos\left(n\omega_xx\right)\cos\left(m\omega_yy\right), \cos\left(n\omega_xx\right)\sin\left(m\omega_yy\right), \sin\left(n\omega_xx\right)\cos\left(m\omega_yy\right), \sin\left(n\omega_xx\right)\sin\left(m\omega_yy\right) \right\}_{m,n=0,1,2,\cdots}$$
with $\omega_x = \frac{2\pi}{P_x}$ and $\omega_y = \frac{2\pi}{P_y}$.
If we only choose $m,n\in \{0,1\}$, the basis are: 
\begin{multline*}
\left\{ 1, \cos(\omega_xx), \sin(\omega_xx), \cos(\omega_yy), \sin(\omega_yy), \right. \\
\left. \cos(\omega_xx)\cos(\omega_yy), \cos(\omega_xx)\sin(\omega_yy), \sin(\omega_xx)\cos(\omega_yy), \sin(\omega_xx)\sin(\omega_yy) \right\}
\end{multline*}
or equivalently (due to trigonometric identities): 
\begin{equation*}
\left\{ 1, \cos\left(\omega_xx - \phi\right), \cos\left(\omega_yy - \phi\right), \cos\left(\omega_xx \pm \omega_yy - \phi\right): \phi \in \{0,\frac{\pi}{2}\} \right\}.
\end{equation*}
Here, $\phi$ is the phase offset and has two values (0 and $\frac{\pi}{2}$). However, these phase offsets may not be optimal, and thus instead of using fixed phase offsets, we can use multiple phase offsets and set them as trainable parameters:
\begin{multline*}
\{ 1, \cos\left(\omega_xx - \phi_{1,1}\right), \cdots, \cos\left(\omega_xx - \phi_{1,m}\right), \cos\left(\omega_yy - \phi_{2,1}\right), \cdots, \cos(\omega_yy-\phi_{2,m}), \\ \cos(\omega_xx + \omega_yy - \phi_{3,1}),\cdots,\cos(\omega_xx + \omega_yy - \phi_{3,m}), \cos(\omega_xx - \omega_yy - \phi_{4,1}),\cdots,\cos(\omega_xx - \omega_yy - \phi_{4,m}) \},
\end{multline*}
where $\phi_{i,j}$ ($i=1,\cdots,4$ and $j=1,\cdots,m$) are trainable.


\subsubsection{POD-DeepONet: Precomputed POD basis as the trunk net}

The vanilla DeepONet uses the trunk net to automatically learn the basis of the output function from the data. Here, we propose the POD-DeepONet, where we compute the basis by performing proper orthogonal decomposition (POD) on the training data (after first removing the mean). Then, we use this POD basis as the trunk net and only use neural networks for the branch net to learn the coefficients of the POD basis (Fig.~\ref{fig:deeponet}A). The output can be written as:
$$\mathcal{G}(v)(\xi) = \sum_{k=1}^p b_k(v) \phi_k(\xi) + \phi_0(\xi),$$
where $\phi_0(\xi)$ is the mean function of $u(\xi)$ computed from the training dataset. $\{b_1, b_2, \dots, b_p\}$ are the $p$ outputs of the branch net, and $\{\phi_1, \phi_2, \dots, \phi_p\}$ are the $p$ precomputed POD modes of $u(\xi)$. The proposed POD-DeepONet shares a similar idea of using POD to represent functions as in \cite{bhattacharya2020model}.

\subsubsection{Rescaling DeepONet with second moment analysis}
\label{sec:deeponet_scale}

It has been shown that it could be difficult to train neural networks if the variance of the output of each layer vanishes or explodes \cite{glorot2010understanding,CiCP-28-1671}. To facilitate network training, it is beneficial to have unit variance after the random initialization of neural networks. Based on this idea, different initialization methods have been developed for different activation functions, e.g., the Glorot initialization for $\tanh$ \cite{glorot2010understanding} and the He initialization for ReLU \cite{he2015delving}.

In DeepONet, the branch and trunk nets are two independent networks, and thus existing methods can be applied directly to maintain its variance. However, the DeepONet output is a product of the outputs of branch net and trunk net, which is not guaranteed to have unit variance. Hence, it is important to rescale the DeepONet output by its standard deviation as:
$$\mathcal{G}(v)(\xi) = \frac{1}{\sqrt{\var{\sum_{k=1}^p b_k(v) t_k(\xi)}}} \left[\sum_{k=1}^p b_k(v) t_k(\xi) + b_0 \right].$$
 
Next, we analyze the variance $\var{\sum_{k=1}^p b_k(v) t_k(\xi)}$. We assume that both the branch net and the trunk net are standard fully-connected neural networks:
\begin{gather*}
    b_k(v(x_1),v(x_2),\ldots,v(x_m)) = w_{k,L_b}^b \sigma_1 ( b^{L_b-1}(v(x_1),v(x_2),\ldots,v(x_m))), \qquad (w_{k,L_b}^b)^\top, b^{L_b-1}\in \Real^{ n_{k,L_b}^b}, \\
	t_k  = \sigma_2 (w_{k,L_{t}-1}^t t^{L_t-1}(x) +\theta_{k}^t),\qquad (w_{k,L_{t}-1}^t)^\top, t^{L_t-1}\in \Real^{ n_{k,L_t-1}^t},
\end{gather*}
where $b^{L_b-1}$ and $t^{L_t-1}$ are the outputs of the last hidden layer of the branch net and the trunk net, respectively, and they are fully-connected ReLU networks.

We consider the ReLU activation function in both the branch and trunk nets, and thus use the He initialization \cite{he2015delving}. 
We assume that all the biases are initialized to 0 and all the weights have the Gaussian distribution with mean $0$ while all the random variables are independent and the weights of the same layer have the same variance.
Then, at the initialization step we have \cite{he2015delving}:
\begin{gather*}
  \mean{b_k}=0,\quad   
  \mean{b_k^2}=\mean{(w_{k,L_b}^b)^2}
  \mean{\sigma_1^2 ( b^{L_b-1})}=
  \frac{1}{2}n_{k,L_b}^b\var{
  (w_{k,L_b}^b)_1}\var{(b^{L_b-1})_1}, \\
    \mean{t_k^2} = \frac{1}{2}
    \mean{(w_{k,L_t-1}^t t^{L_t-1})^2}
    =\frac{1}{4}n_{k,L_t-1}^t \var{(w_{k,L_t-1}^t)_1}\var{
    (t^{L_t-1})_1}.
\end{gather*}
As in the He initialization, we initialize the weights such that  
$\var{b^{(L_b-1})_1} =1/2$ and  $\var{(t^{L_t-1})_1} =1/2$.  Taking $\frac{1}{2}n_{k,L_b}^b\var{(w_{k,L_b}^b)_1}=1$ and 
$\frac{1}{2}n_{k,L_t-1}^t \var{(w_{k,L_t-1}^t)_1}=1$, we then have 
\begin{equation*}
 \var{b_k}  = \frac{1}{2},\quad   \mean{t_k^2} = \frac{1}{4}.
\end{equation*}
%
Observe that  $b_k$ and $t_k$ are independent and that $\mathbb{E}\lbrack  b_k \rbrack =0$ and $\mean{b_kt_k} =0$.  Moreover,
 $\mean{b_ib_j } =1/2$ if $i=j$ and $\mean{b_ib_j } =0$ if $i\neq j$.  
Then 
\begin{eqnarray*}
\var{\sum_{k=1}^p b_k t_k } & =& \mean{(\sum_{k=1}^p b_k t_k)^2 } =\mean{\sum_{k,l=1}^p b_k b_l t_kt_l }  =\sum_{k=1}^p   \mathbb{E}\lbrack  b_k^2 \rbrack  \mathbb{E}\lbrack  t_k^2 \rbrack =
\sum_{k=1}^p 
\var{b_k} \mean{  t_k ^2 } =\frac{p}{8}.
\end{eqnarray*} 

The analysis above suggests that the scaling factor of the DeepONet should be $\mathcal{O}(1/\sqrt{p})$. 
In our  experiment with POD-DeepONet, we indeed obtain better accuracy when it is scaled by $\mathcal{O}(1/\sqrt{p})$. 
However, the analysis applies to the He initialization and only considers the initialization stage, thus the scaling $\mathcal{O}(1/\sqrt{p})$ may not be optimal during network training in practice.
In some numerical experiments, we find that the vanilla DeepONet works well even without scaling. POD-DeepONet obtains a bit better accuracy when it is scaled by  $\mathcal{O}(1/p)$.


\subsubsection{Multiple outputs}

We have discussed the DeepONet for a single output function, but this can be extended to multiple output functions. Let us assume that we have $n$ output functions. Here we propose a few possible approaches:
\begin{enumerate}
    \item The simplest approach is that we can directly use $n$ independent DeepONets, and each DeepONet outputs only one function.
    \item The second approach is that we can split the outputs of both the branch net and the trunk net into $n$ groups, and then the $k$th group outputs the $k$th solution. For example, if $n=2$ and both the branch and trunk nets have 100 output neurons, then the dot product between the first 50 neurons of the branch and trunk nets generates the first function, and the remaining 50 neurons generate the second function.
    \item The third approach is similar to the second approach, but we only split the branch net and share the trunk net. For example, for $n=2$, we split the 100 output neurons of the branch net into 2 groups, but the trunk net only has 50 output neurons. Then, we use the first group of the branch net and the entire trunk net to generate the first output, and use the second group and the trunk net to generate the second output.
    \item Similarly, we can also split the trunk net and share the branch net.
\end{enumerate}

It is not easy to determine a priori which approach would work best, as this may be highly problem dependent. In this work, we use the second approach. More investigation and comparison between these four approaches both theoretically and computationally should be done in the future.

\subsubsection{Fast algorithms and code implementation}

For different types of data sets, we can use different code implementation for DeepONet to dramatically reduce the computational cost and memory usage by orders of magnitudes. We consider the following three cases:
\begin{enumerate}
    \item \textbf{The straightforward way (flexible but expensive).} If different output $u$ has different number and different locations of $\xi$, then each data point is a triplet $(v, \xi, u(\xi))$. This implementation can be used for any problem setup.
    \item \textbf{Removing the redundancy in branch net input (cheaper).} If the number of trunk net input $\xi$ is the same for all branch net input $v$ (the values of $\xi$ can be different for different $v$), then we can share the same input $v$ in the branch net for all the corresponding $\xi$.
    \item \textbf{Removing the redundancy in both branch and trunk net inputs (cheapest).} In addition to Case 2 above, if the trunk net input $\xi$ is the same for all branch net input $v$, then all input $v$ can share the same trunk net input $\xi$.
\end{enumerate}
Here, we only introduce the underlying idea, but for the detailed code implementation, we refer the reader to our code.

\subsection{FNO}

Next, we introduce the vanilla FNO and then discuss how to generalize the vanilla FNO for problems whose inputs and outputs are defined on different domains or on a complex geometry.

\subsubsection{Vanilla FNO}
\label{sec:fno}

FNO is formulated by parameterizing the integral kernel directly in Fourier space. The main network parameters are defined and learned in the Fourier space rather than the physical space, i.e., the coefficients of the Fourier series of the output function are learned from the data.

Different from DeepONet, FNO discretizes both the input function $v(x)$ and the output function $u(\xi)$ by using pointwise evaluations in an equispaced mesh. Also, FNO requires that $v$ and $u$ are defined on the domain (i.e., $D = D'$, and $x$ and $\xi$ become the same variable), and the same discretization (i.e., the same mesh) is used for $v$ and $u$. In short, FNO maps the discretization of $v$ in an equispaced mesh to the discretization of $u$ on the same mesh. For example, if $D$ is a rectangular domain in 2D, then FNO input $v$ is a 2D matrix (i.e., an image). When we present the architecture of FNO, we will discuss both the function value on one mesh location and also the function values on all mesh locations, so we use the following convention for a function $f$: $f(x)$ means the value of $f(x)$ in a mesh location $x$, and the bold font $\bm{f}$ means the values of $f(x)$ in all mesh locations.

\begin{figure}[htbp]
    \centering
    \includegraphics[width=\textwidth]{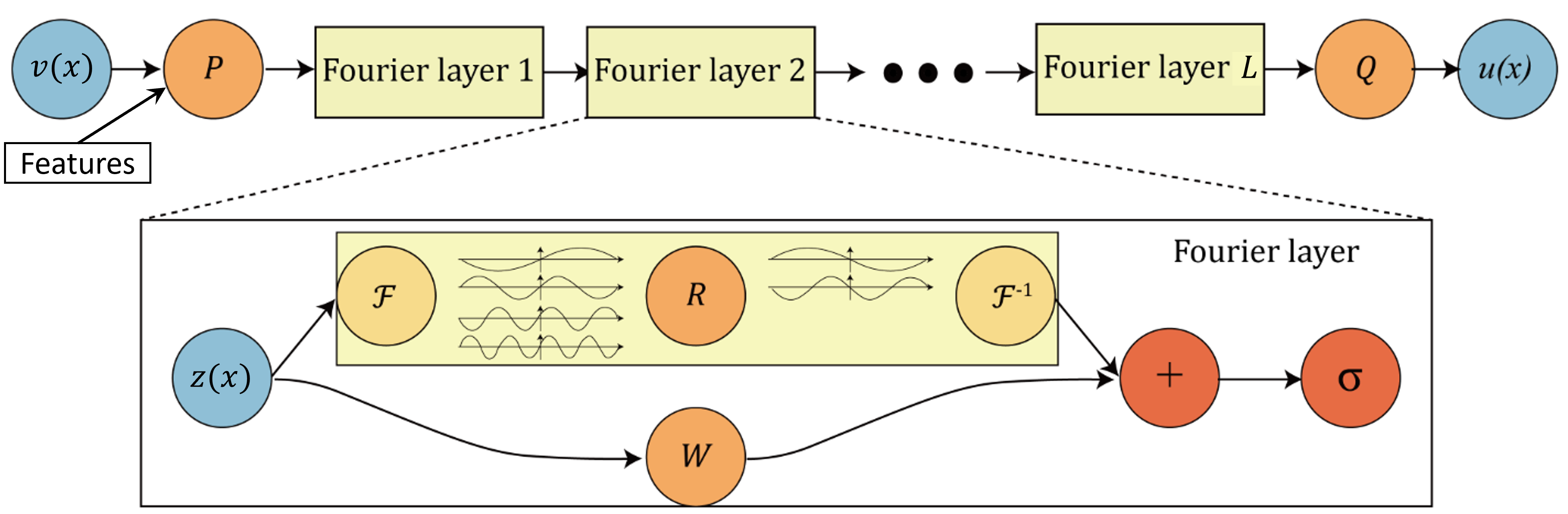}
    \caption{\textbf{Architecture of FNO with extra features added to the input.} Figure is adapted from \cite{li2020fourier}.}
    \label{fig:fno}
\end{figure}

In FNO, for any location $x$ on the mesh, the function value $v(x)$ is first lifted to a higher dimensional representation $z_0(x)$ by
$$z_0(x) = P(v(x)) \in \mathbb{R}^{d_z}$$
using a local transformation $P: \mathbb{R} \to \mathbb{R}^{d_z}$ (Fig.~\ref{fig:fno}), which is parameterized by a shallow fully-connected neural network or simply a linear layer. We note that $\bm{z}_0$ is defined on the same mesh as $\bm{v}$, and the values of $\bm{z}_0$ in the mesh can be viewed as an image with $d_z$ channels. Then $L$ Fourier layers are applied iteratively to $\bm{z}_0$. Let us denote $\bm{z}_L$ as the output of the last Fourier layer, and the dimension of $z_L(x)$ is also $d_z$. Hence, at the end, another local transformation $Q: \mathbb{R}^{d_z} \to \mathbb{R}$ is applied to project $z_L(x)$ to the output (Fig.~\ref{fig:fno}) by
$$u(x) = Q(z_L(x)).$$
We parameterize $Q$ by a fully-connected neural network.

Next we introduce the Fourier layer by using the Fast Fourier Transform (FFT). For the output of the $l$th Fourier layer $\bm{z}_{l}$ with $d_v$ channels, we first compute the following transform by FFT $\mathcal{F}$ and inverse FFT $\mathcal{F}^{-1}$ (the top path of the Fourier layer in Fig.~\ref{fig:fno}):
$$\mathcal{F}^{-1} \left( R_l \cdot \mathcal{F}(\bm{z}_l) \right).$$
The details are as follows:
\begin{itemize}
    \item $\mathcal{F}$ is applied to each channel of $\bm{z}_l$ separately, and we usually truncate the higher modes of $\mathcal{F}(\bm{z}_l)$, keeping only the first $k$ Fourier modes in each channel. So $\mathcal{F}(\bm{z}_l)$ has the shape $d_v \times k$. For 2D functions, $k = k_1 \times k_2$, where $k_1$ and $k_2$ are the number of modes to keep in the first and second dimensions, respectively. For 3D functions, it can be done similarly.
    \item We apply a different (complex-number) weight matrix of shape $d_v \times d_v$ for each mode index of $\mathcal{F}(\bm{z}_l)$, so we have $k$ trainable matrices, which form a weight tensor $R_l \in \mathbb{C}^{d_v \times d_v \times k}$. Then, $R_l \cdot \mathcal{F}(\bm{z}_l)$ has the same shape of $d_v \times k$ as $\mathcal{F}(\bm{z}_l)$.
    \item Before we perform the inverse FFT, we need to append zeros to $R_l \cdot \mathcal{F}(\bm{z}_l)$ to fill in the truncated modes.
\end{itemize}
Moreover, in each Fourier layer, a residual connection with a weight matrix $W_l \in \mathbb{R}^{d_v \times d_v}$ is used to compute a new set of $d_v$ channels, and each new channel is a linear combinations between all the $\bm{z}_l$ channels (the bottom path of the Fourier layer in Fig.~\ref{fig:fno}). $W_l \cdot \bm{z}_l$ has the same shape as $\bm{z}_l$. We can implement $W_l \cdot \bm{z}_l$ by a matrix multiplication or by a convolution layer with the kernel size 1. Then, the output of the $(l+1)$th Fourier layer $\bm{z}_{l+1}$ is
\begin{equation*}
\bm{z}_{l+1} =
\sigma \left(
\mathcal{F}^{-1} \left( R_l \cdot \mathcal{F}(\bm{z}_l) \right) + W_l \cdot \bm{z}_l + \bm{b}_l
\right),
\end{equation*}
where $\sigma$ is a nonlinear activation function, and $\bm{b}_l \in \mathbb{R}^{d_v}$ is a bias. In this work, we use ReLU in all cases.

We note that in practice, to achieve good accuracy, we need to use the coordinates $x$ as the input as well, in addition to $v(x)$, i.e., the FNO input is the values of $(x, v(x)) \in \mathbb{R}^{d+1}$ in a grid (i.e., an image with $d+1$ channels). This is a special case of the feature expansion that we will discuss in Section~\ref{sec:fno-feature}.

\subsubsection{dFNO+: Operators with inputs and outputs defined on different domains}
\label{sec:FNO_domains}

One limitation of FNO is that it requires $D$ and $D'$ to be the same domain, which is not always satisfied. 
Here, we discuss two scenarios where $D \neq D'$, and propose new extensions of FNO to address this issue.

\paragraph{Case I: The output space is a product space of the input space and another space $D_0$, i.e., $D' = D \times D_0$} We use a specific example to illustrate the idea. We consider the PDE solution operator mapping from the initial condition to the solution in the whole domain:
$$\mathcal{G}: v(x) = u(x, 0) \mapsto u(x,t),$$
where $x \in [0, 1]$ and $t \in [0, T]$. Here, $D = [0, 1]$, $D' = [0, 1] \times [0, T]$, and thus $D_0 = [0, T]$. In order to match the input and output domains, we propose the following two methods.

\textbf{Method 1: Expand the input domain.} We can extend the input function by adding the extra coordinate $t$, defining $\tilde{v}$ as
$$\tilde{v}(x, t) = v(x).$$
Then, FNO is used to learn the mapping from $\tilde{v}(x, t)$ to $u(x, t)$.

\textbf{Method 2 \cite{li2020fourier}: Shrink the output domain via RNN.} We can also reduce the dimension of the output by decomposing $\mathcal{G}$ into a series of operators. We denote a new time-marching operator
\begin{equation*}
    \tilde{\mathcal{G}}: u(x, t) \mapsto u(x, t + \Delta t), \quad x \in D,
\end{equation*}
i.e, $\tilde{\mathcal{G}}$ predicts the solution at $t + \Delta t$ from the solution at $t$. Then, we apply $\tilde{\mathcal{G}}$ to the input $v(x)$ repeatedly to obtain the solution in the whole domain, which is similar to a RNN.

\paragraph{Case II: The input space is a subset of the output space, i.e., $D \subset D'$} In general, we can always extend $v$ from $D$ to $D'$ by padding zeros in the domain $D'\setminus D$, i.e., we define
$$\tilde{v}(\xi) = \begin{cases}
v(\xi), & \text{if } \xi \in D \\
0, & \text{if } \xi \in D'\setminus D
\end{cases},
$$
and then learn the mapping from $\tilde{v}$ to $u$.

However, in some cases, this padding strategy may not be efficient. For example, we consider a PDE defined on a rectangular domain $(x,y) \in D'= [0,1]^2$, and the operator is the mapping from the Dirichlet boundary condition $v$ defined in the four boundaries ($D = \{0, 1\}\times[0, 1] \cup [0,1]\times\{0, 1\}$) to the solution $u(x,y)$ inside the rectangular domain. In this example, $D$ is essentially a 1D space and occupies zero area in $D'$. We propose a better strategy so that we first unfold the curve of $v$ into a 1D function $\tilde{v}$ defined in $[0, 4]$:
$$\tilde{v}(\tilde{x}) = \begin{cases}
v(\tilde{x}, 0), & \text{if } \tilde{x} \in [0, 1] \text{ (bottom boundary)} \\
v(1, \tilde{x} - 1), & \text{if } \tilde{x} \in [1, 2] \text{ (right boundary)} \\
v(3 - \tilde{x}, 1), & \text{if } \tilde{x} \in [2, 3] \text{ (top boundary)} \\
v(0, 4 - \tilde{x}), & \text{if } \tilde{x} \in [3, 4] \text{ (left boundary)} \\
\end{cases},
$$
and then, we use the method in Case I above to learn the operator from $\tilde{v}$ in 1D to $u$ in 2D.

\subsubsection{gFNO+: Operators with inputs and outputs defined on a complex geometry}
\label{sec:gfno}

FNO uses FFT, which requires the input and output functions to be defined on a Cartesian domain with a lattice grid mesh. However, for the PDEs defined on a complex geometry $D$ (e.g., L-shape, triangular domain, etc), an unstructured mesh is usually used, and thus we need to deal with two issues: (1) non-Cartesian domain, and (2) non-lattice mesh. For the second issue of unstructured mesh, we need to do interpolation between the unstructured mesh and a lattice grid mesh.

For the issue of the Cartesian domain, we first define the Cartesian domain $\tilde{D}$, which is the minimum bounding box of $D$, and then extend $v$ (the same for $u$) from $D$ to $\tilde{D}$ by
$$\tilde{v}(x) = \begin{cases}
v(x), & \text{if } x \in D \\
v_0(x), & \text{if } x \in \tilde{D} \setminus D
\end{cases}.
$$
Here, the choice of $v_0(x)$ is not unique. The simplest choice is $v_0(x) = 0$, i.e., zero padding. However, we find that such zero padding leads to large error of FNO, which may be due to the discontinuity from $v(x)$ to $v_0(x)$. We propose to compute $v_0(x)$ by ``nearest neighbor'':
$$\text{for } x \in \tilde{D} \setminus D, \quad v_0(x) = v(x_0), \quad \text{where } x_0 = \min_{p \in D} \| p - x\|,$$
so that $\tilde{v}(x)$ is continuous on the boundary of $D$. In the training, we use a mask to only consider the points inside $D$ in the loss function.

\subsubsection{Feature expansion}
\label{sec:fno-feature}
In FNO, the features can be applied by using them as extra network inputs. As we mentioned at the end of Section~\ref{sec:fno}, in practice we need to use the coordinates $x$ as the input, which is a special case of feature expansion. If we have a feature $f(x)$, then the FNO input is the values of $(x, v(x), f(x)) \in \mathbb{R}^{d+2}$, i.e., an image with $d+2$ channels.

\subsection{Comparison between DeepONet and FNO}

Here, we list a comparison between DeepONet and FNO in Table~\ref{tab:deeponet_fno} in terms of some of their properties instead of accuracy. The first three points have been discussed in the introduction above of DeepONet and FNO. Because FNO also discretizes the output function, then after the network training, it can only predict the solution in the same mesh as the input function, whereas DeepONet can make predictions at any location. For the training, FNO requires a full field observation data, but DeepONet is more flexible, except that POD-DeepONet requires a full field data to compute the POD modes.

\begin{table}[htbp]
\centering
\begin{tabular}{c|cc}
\toprule
 & DeepONet & FNO \\
\midrule
Input domain $D'$ \& Output domain $D'$ & Arbitrary & Cuboid, $D=D'$ \\
Discretization of output function $u$ & No & Yes \\ 
Mesh & Arbitrary & Grid \\
Prediction location & Arbitrary & Grid points \\
Full field observation data & No & Yes \\
\bottomrule
\end{tabular}
\caption{\textbf{Comparison between vanilla DeepONet and vanilla FNO.}}
\label{tab:deeponet_fno}
\end{table}

\subsection{Other technical details}

There are several other useful techniques that may improve the performance of DeepONet and FNO, such as learning rate decay, $L^2$ regularization, and input normalization. Here, we emphasize the output normalization. Let us assume that in the training dataset the mean function and the standard deviation function of $u$ is $\bar{u}(\xi)$ and $\text{std}[u](\xi)$, respectively. Then, we construct the surrogate model as
$$u(\xi) = \text{std}[u](\xi) \cdot \mathcal{N}(v)(\xi) + \bar{u}(\xi),$$
where $\mathcal{N}$ is a DeepONet or FNO. Hence, for any $\xi$, the mean value of 
$\mathcal{N}(v)(\xi)$ 
is zero and the standard deviation is one.




\input{theory}

\section{Numerical results}
\label{sec:results}

We compare the performance of DeepONet and FNO on 16 different problems listed in Table~\ref{tab:problems}. To evaluate the performance of the networks, we compute the $L^2$ relative error of the predictions, and for each case, five independent training trials are performed to compute the mean error and the standard deviation. The dataset sizes for each problem are listed in Table~\ref{tab:data_size}, and the network sizes of DeepONets are listed in Tables~\ref{tab:DeepONet_architecture} and \ref{tab:POD_DeepONet_architecture}.

\begin{table}[htbp]
    \centering
    \begin{tabular}{l|l}
    \toprule
    Section & Problems \\
    \midrule
    \cref{sec:burgers} & Burgers' equation \\
    \cref{sec:darcy} & 5 Darcy problems in a rectangular domain and complex geometries \\
    \cref{sec:electroconvection} & Multiphysics electroconvection problem \\
    \cref{sec:advection} & 3 Advection problems \\
    \cref{sec:instability_wave} & Linear instability waves in high-speed boundary layers \\
    \cref{sec:euler} & Compressible Euler equation with non-equilibrium chemistry \\
    \cref{sec:flapping_airfoil} & Predicting surface vorticity of a flapping airfoil \\
    \cref{sec:ns} & Navier-Stokes equation in the vorticity-velocity form \\
    \cref{sec:cavity} & 2 problems of regularized cavity flows \\
    \bottomrule
    \end{tabular}
    \caption{\textbf{16 problems tested in this study.}}
    \label{tab:problems}
\end{table}

\subsection{Burgers' equation}
\label{sec:burgers}

\paragraph{Problem setup}

We first consider the one-dimensional Burgers' equation:
\begin{equation*}\label{eq:Burgers}
    \frac{\partial u}{\partial t}+ u \frac{\partial u}{\partial x} = \nu  \frac{\partial^{2} u}{\partial  x^{2} }, \quad x\in (0,1), ~ t\in (0,1],
\end{equation*}
with periodic boundary condition, where $\nu=0.1$ is the viscosity. Here, we learn the operator mapping from the initial condition $u(x,0) = u_{0}(x)$ to the solution $u(x,t)$ at $t=1$, i.e., $$\mathcal{G}: u_{0}(x) \mapsto u(x,1).$$
We use the dataset generated in Ref.~\cite{li2020fourier}, where the initial condition is generated according to a random distribution $\mu=\mathcal{N}(0, 625(-\Delta+25I)^{-2})$. We use a spatial resolution with 128 grids to represent the input and output functions.

\paragraph{Results}

As discussed in Section~\ref{sec:deeponet-bc}, in DeepONet, we impose
the periodic boundary condition by applying four Fourier basis $\{\cos(2\pi x), \sin(2\pi x), \cos(4\pi x), \sin(4\pi x)\}$ to the input of the trunk net. By using the output normalization, the error of DeepONet decreases slightly from 2.29$\pm$0.10\% to 2.15$\pm$0.09\% (Table~\ref{tab:burgers}), and thus DeepONet is not sensitive to the output normalization. FNO is more sensitive to the output normalization, and by using the output normalization FNO achieves an error of 1.93$\pm$0.04\% (Table~\ref{tab:burgers}).

As we showed in Section~\ref{sec:deeponet_scale}, DeepONet has variance proportional to $p$ instead of 1 when each branch and trunk net has variance 1 and thus a scale of $1/\sqrt{p}$ is needed. A different scaling is required for POD-DeepONet as the precomputed POD modes are in place of the trunk nets and are not networks to be trained. We tested different scaling factors, including $1/\sqrt{p}$, $1/p$ and $1/p^{1.5}$ (Table~\ref{tab:burgers}), and POD-DeepONet with the factor $1/p$ has the smallest error of 1.94$\pm$0.07\%, which is almost the same as the error of FNO with the output normalization. For the following problems all POD-deepONet, we use the scale $1/p$.

\begin{table}[htbp]
\centering
\begin{tabular}{c|c}
\toprule
 & \cref{sec:burgers} Burgers' \\
\midrule
DeepONet w/o normalization & 2.29$\pm$0.10\% \\
DeepONet w/ normalization & 2.15$\pm$0.09\% \\
FNO w/o normalization & 2.23$\pm$0.04\% \\
FNO w/ normalization & \textbf{1.93$\pm$0.04}\% \\
\hdashline
POD-DeepONet (w/o rescaling) & 3.46$\pm$0.06\% \\
POD-DeepONet (rescaling by $1/\sqrt{p}$) & 2.40$\pm$0.06\% \\
POD-DeepONet (rescaling by $1/p$) & \textbf{1.94$\pm$0.07}\% \\
POD-DeepONet (rescaling by $1/p^{1.5}$) & 2.41$\pm$0.04\% \\
\bottomrule
\end{tabular}
\caption{\textbf{$L^2$ relative error for the Burgers' equation in \cref{sec:burgers}.} Here, $p$ is the number of outputs of the branch and trunk nets.}
\label{tab:burgers}
\end{table}

\subsection{Darcy problem}
\label{sec:darcy}

We consider two-dimensional Darcy flows in different geometries filled with porous media, which can be described by the following equation:
\begin{equation}\label{eq:darcy}
    -\nabla \cdot (K(x,y) \nabla h(x,y)) = f,  \,(x,y)\in \Omega,
\end{equation}
where $K$ is the permeability field, $h$ is the pressure, and $f$ is a source term which can be either a constant or a space-dependent function. Boundary conditions will be described in the problem setup below. Four different geometries are considered in the present study, including a rectangular domain in Section~\ref{sec:darcy_rectangular}, a pentagram with a hole in Section~\ref{sec:darcy_pentagram}, a triangular domain in Section~\ref{sec:darcy_triangular}, and a triangular domain with notch in Section~\ref{sec:darcy_notch}. We generate the dataset by solving Eq.~\eqref{eq:darcy} using the MATLAB Partial Differential Equation Toolbox (for more details see Section~\ref{sec:darcy_data}).

\subsubsection{Darcy problem in a rectangular domain}
\label{sec:darcy_rectangular}

\paragraph{Problem setup}

The first example of Darcy flow is defined in a rectangular domain $[0,1]^2$ with zero Dirichlet boundary condition. We are interested in learning the mapping from the permeability field $K(x,y)$ to the pressure field $h(x,y)$, i.e.,
$$\mathcal{G}: K(x,y) \mapsto h(x,y).$$
Here, we consider two datasets of different types of permeability fields:
\begin{itemize}
    \item \textbf{Piecewise constant (PWC).} The first dataset is from Ref.~\cite{li2020fourier}. The coefficient field $K$ is defined as $K=\psi(\mu)$, where $\mu=\mathcal{N}(0,(-\Delta+9I)^{-2})$ with zero Neumann boundary conditions on the Laplacian, and the mapping $\psi$ performs the binarization on the function, namely it converts the positive values to 12 and the negative values to 3. The grid resolution of $K$ and $h$ is $29\times 29$.
    \item \textbf{Continuous (Cont.).} We use a truncated Karhunen-Lo\`eve (KL) expansion to express the permeability field $K(x,y) = \exp(F(x,y))$, where $F(x,y)$ denotes a truncated KL expansion for a given Gaussian process. Specifically, we keep the leading 100 terms in the KL expansion for the Gaussian process with zero mean and the following covariance kernel:
    \begin{equation*}
        \mathcal K ((x,y), (x',y')) = \exp \left[\frac{-(x-x')}{2l_1^2} + \frac{-(y-y')^2}{2l_2^2}\right],
    \end{equation*}
    with $l_1 = l_2 = 0.25$. Both $K(\bm{x})$ and $h(\bm{x})$ have the same resolution of $20 \times 20$.
\end{itemize}
Examples of these two Darcy datasets can be found in Fig.~\ref{fig:darcy_testing}.

\paragraph{Results}

We enforce the zero Dirichlet boundary condition on DeepONet by choosing the surrogate solution as
$$\hat{u}(x,y) = 20x(1-x)y(1-y)\mathcal{N}(x,y,K),$$
where $\mathcal{N}$ is a DeepONet, as we discussed in Section~\ref{sec:deeponet-bc}. We use the coefficient 20 such that $20x(1-x)y(1-y)$ is of order 1 for $x \in [0,1]$ and $y\in [0,1]$. Similar to the Burgers' problem, by using the output normalization, a better accuracy of DeepONet and FNO is obtained (Table~\ref{tab:Darcy_rect}). For the case PWC, all the methods have errors around 2\%, and POD-DeepONet achieves the smallest error. For the case Cont., all the methods have the error around 1\%, and FNO is slightly better. However, POD-DeepONet is only worse by one standard deviation of the error (Table~\ref{tab:Darcy_rect}), so there is no significant difference between the performance of POD-DeepONet and FNO.

\begin{table}[htbp]
\centering
\begin{tabular}{c|cc}
\toprule
& \cref{sec:darcy_rectangular} Darcy (PWC) & \cref{sec:darcy_rectangular} Darcy (Cont.) \\
\midrule
DeepONet w/o normalization & 2.91$\pm$0.04\% &  2.04$\pm$0.13\% \\
DeepONet w/ normalization & 2.98$\pm$0.03\% & 1.36$\pm$0.12\% \\
FNO w/o normalization &  4.83$\pm$0.12\%  & 2.38$\pm$0.02\% \\
FNO w/ normalization & 2.41$\pm$0.03\% & \textbf{1.19$\pm$0.05}\% \\
\hdashline
POD-DeepONet & \textbf{2.32$\pm$0.03}\% & \textbf{1.26$\pm$0.07\%} \\
\bottomrule
\end{tabular}
\caption{\textbf{$L^2$ relative error for the Darcy problem in a rectangular domain in \cref{sec:darcy_rectangular}.} PWC, piecewise constant. Cont, continuous.}
\label{tab:Darcy_rect}
\end{table}






\subsubsection{Darcy problem in a pentagram with a hole}
\label{sec:darcy_pentagram}

\begin{figure}[htbp]
\centering
\subfigure[]{
\includegraphics[width=0.24\textwidth]{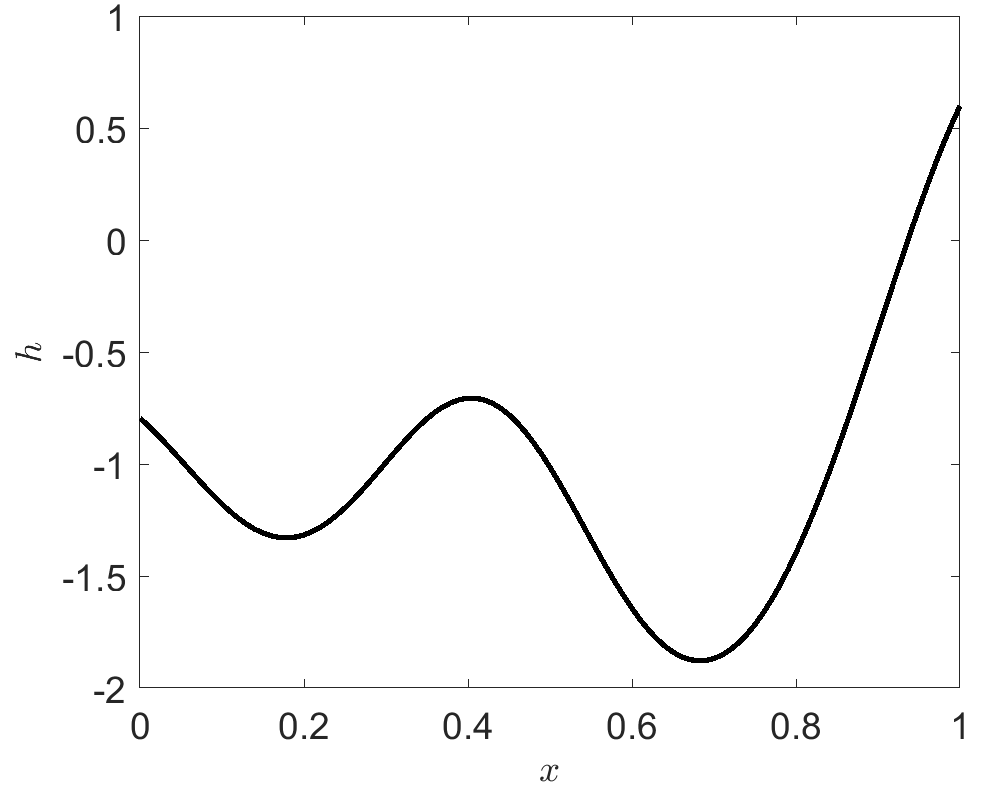}
\includegraphics[width=0.24\textwidth]{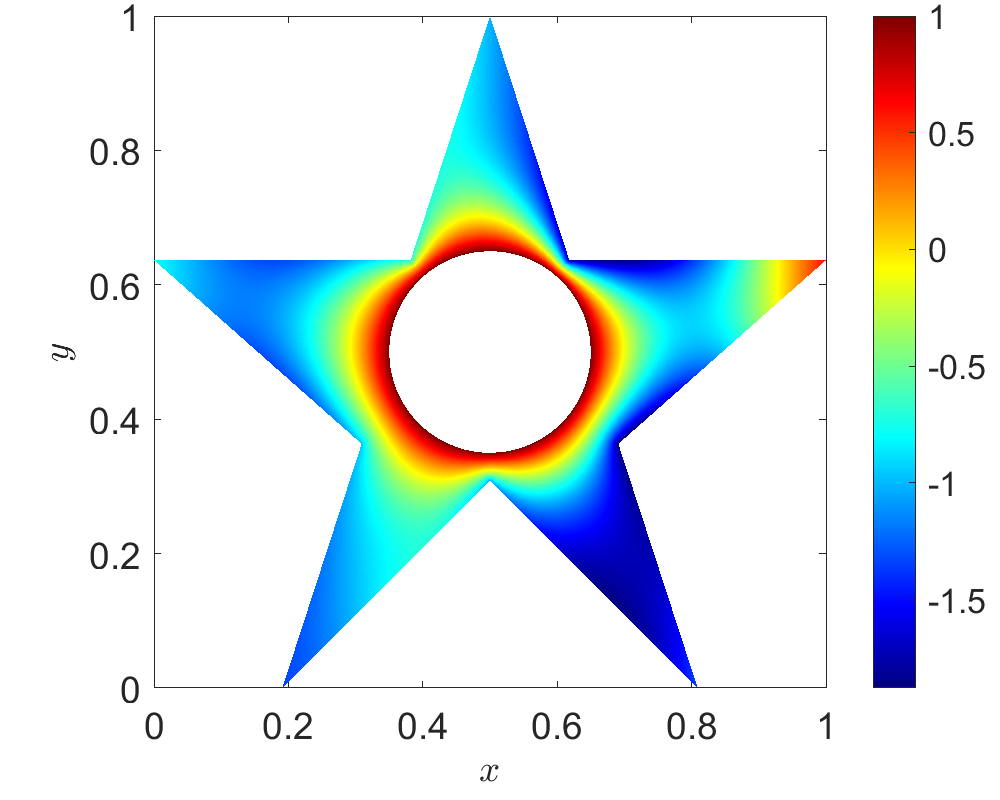}
\includegraphics[width=0.24\textwidth]{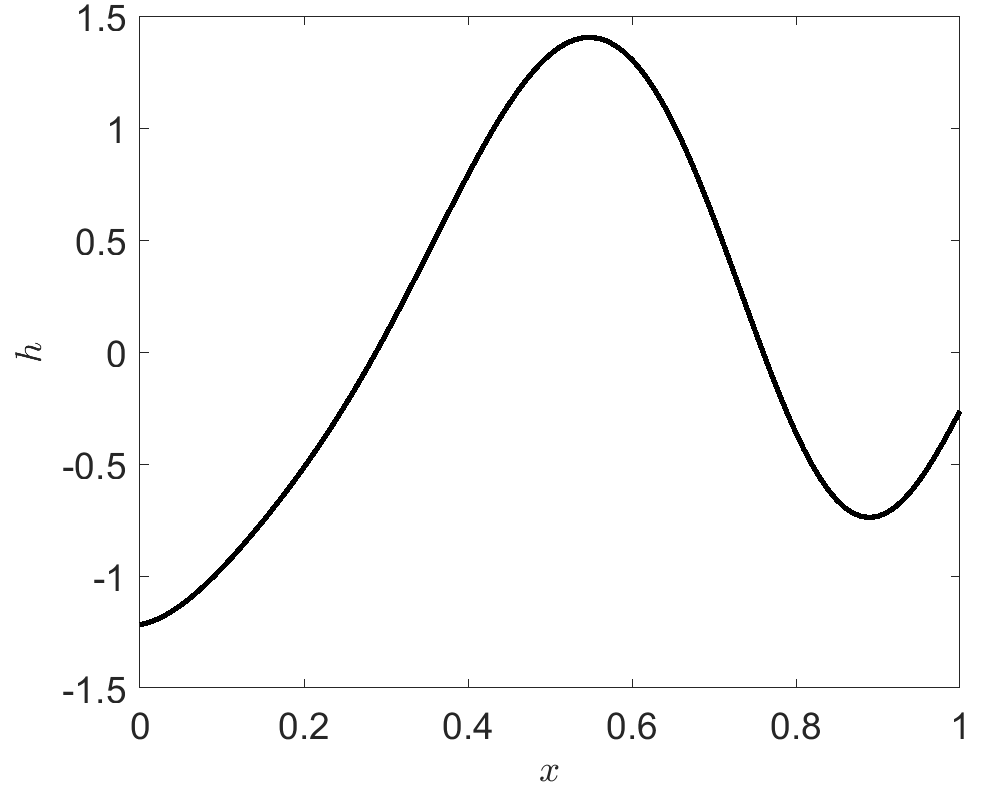}
\includegraphics[width=0.24\textwidth]{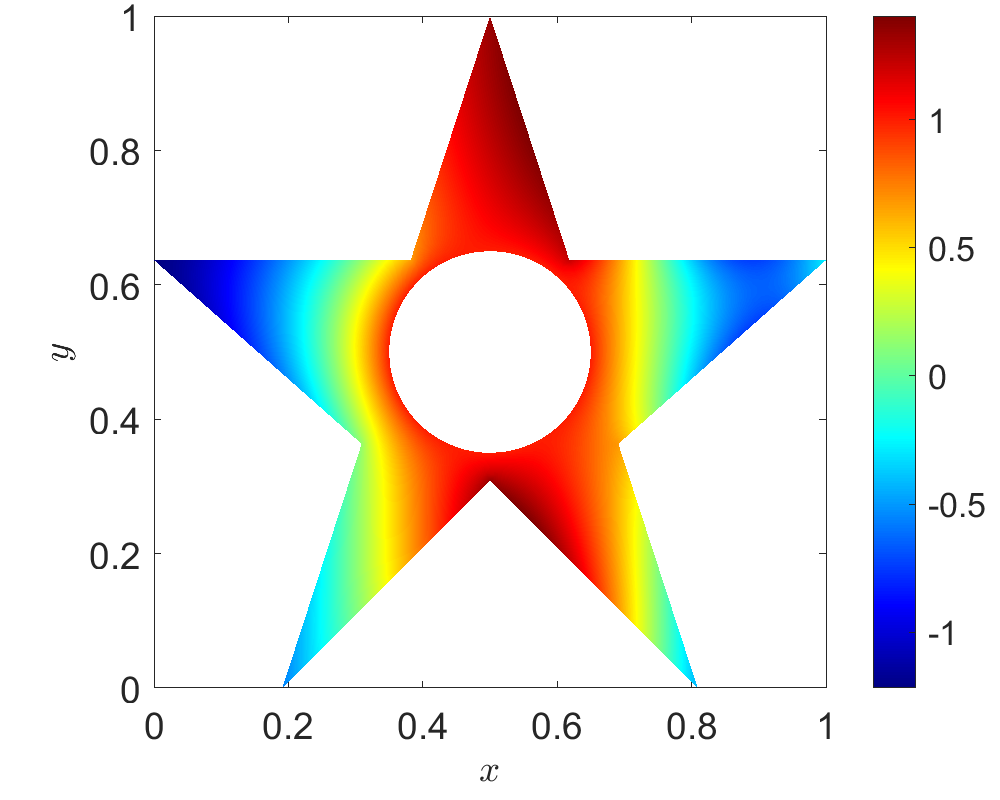}}
\subfigure[]{\includegraphics[width=0.24\textwidth]{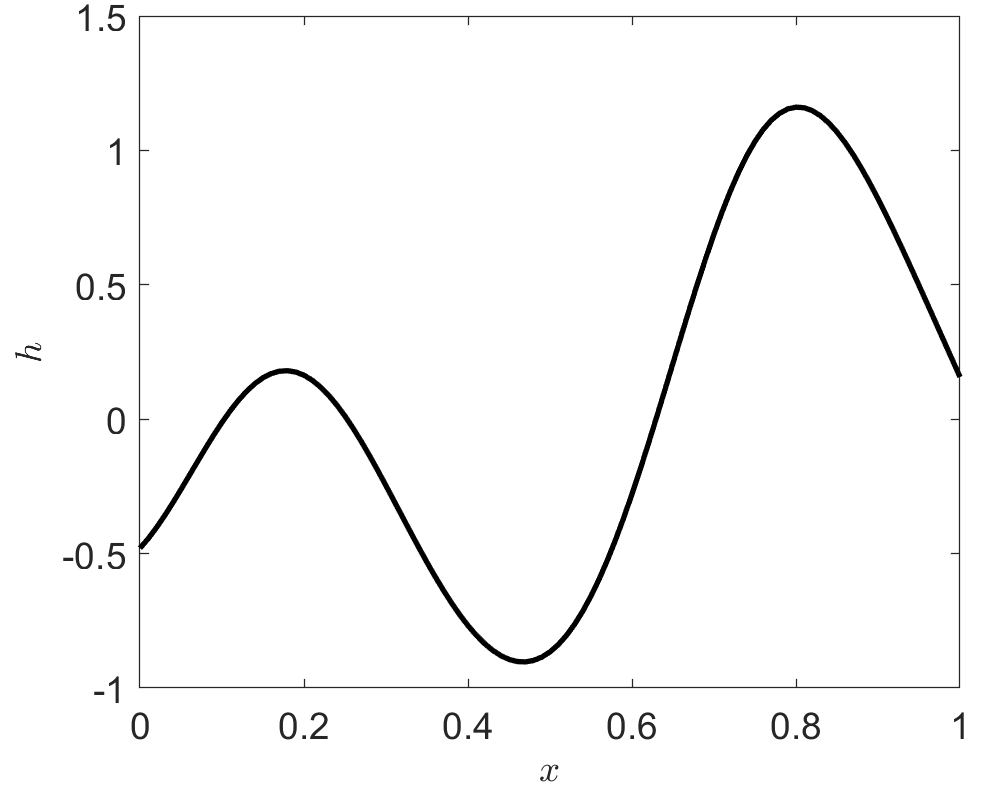}
\includegraphics[width=0.24\textwidth]{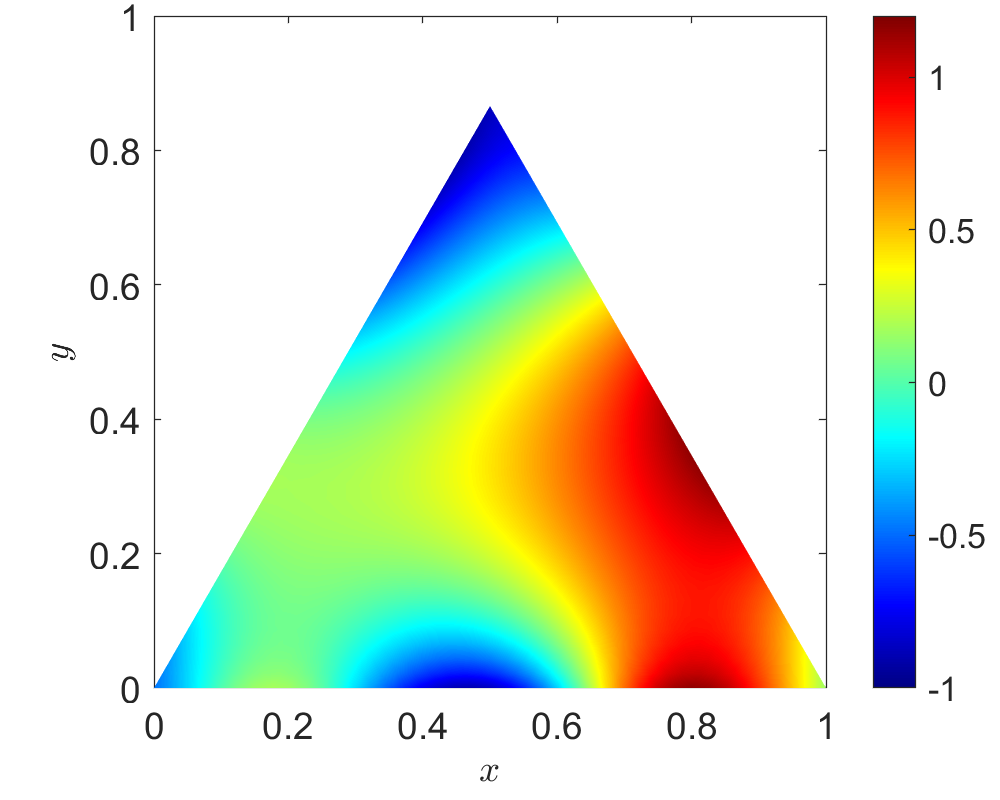}
\includegraphics[width=0.24\textwidth]{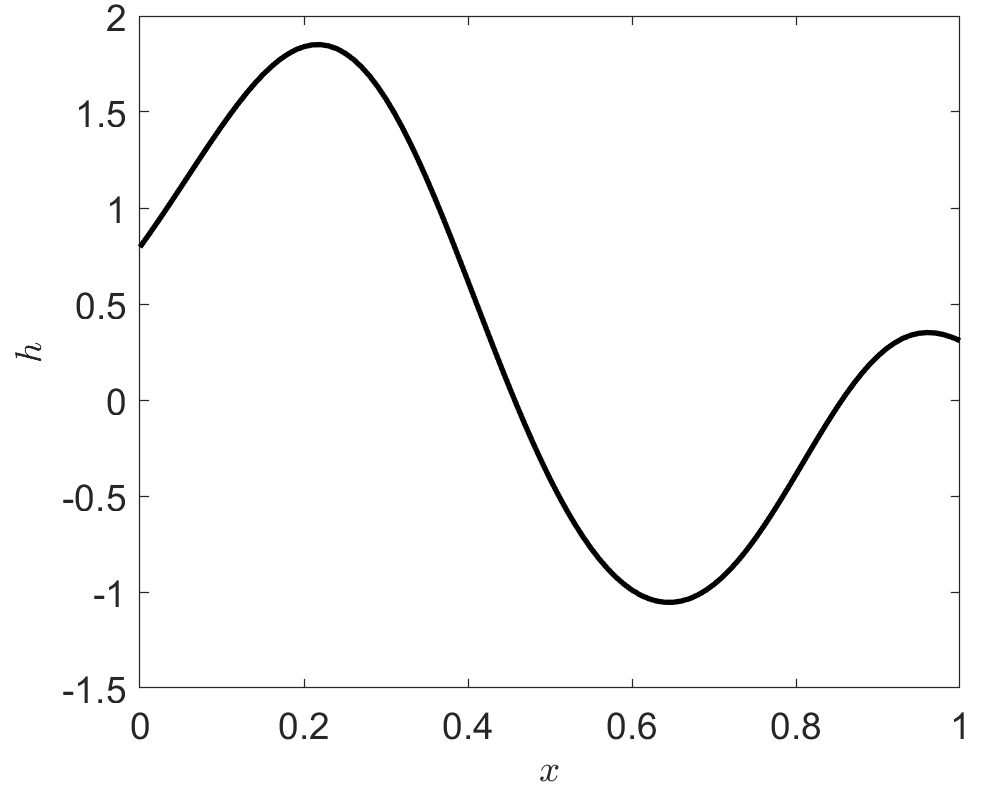}
\includegraphics[width=0.24\textwidth]{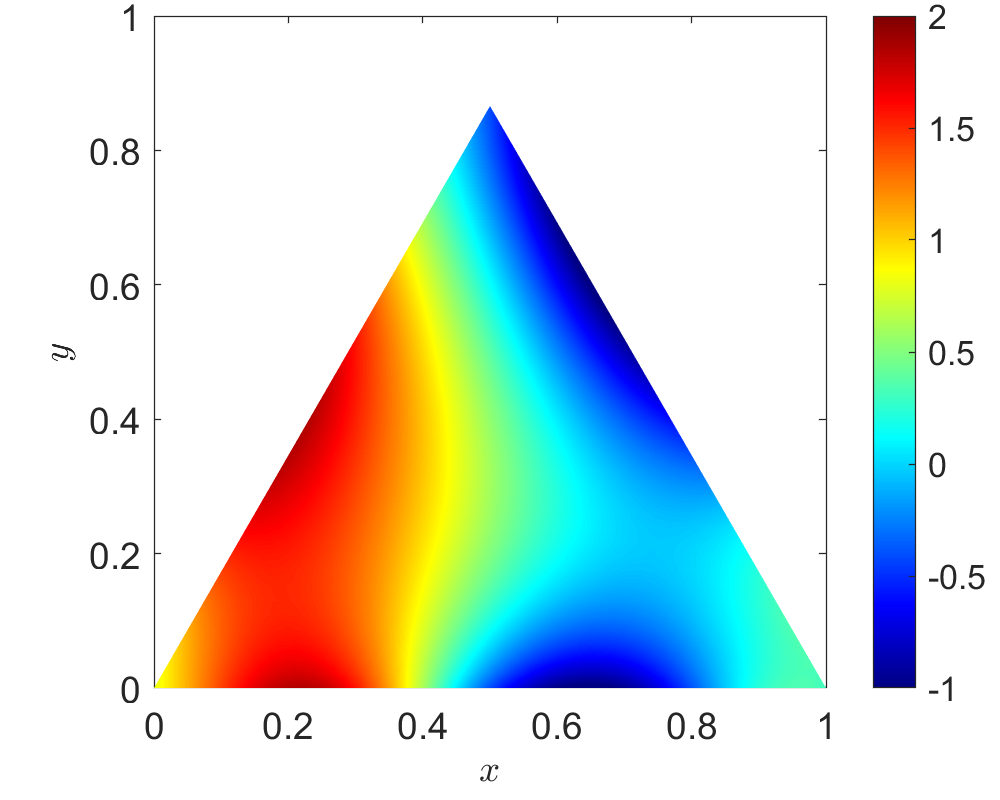}}
\subfigure[]{\includegraphics[width=\textwidth]{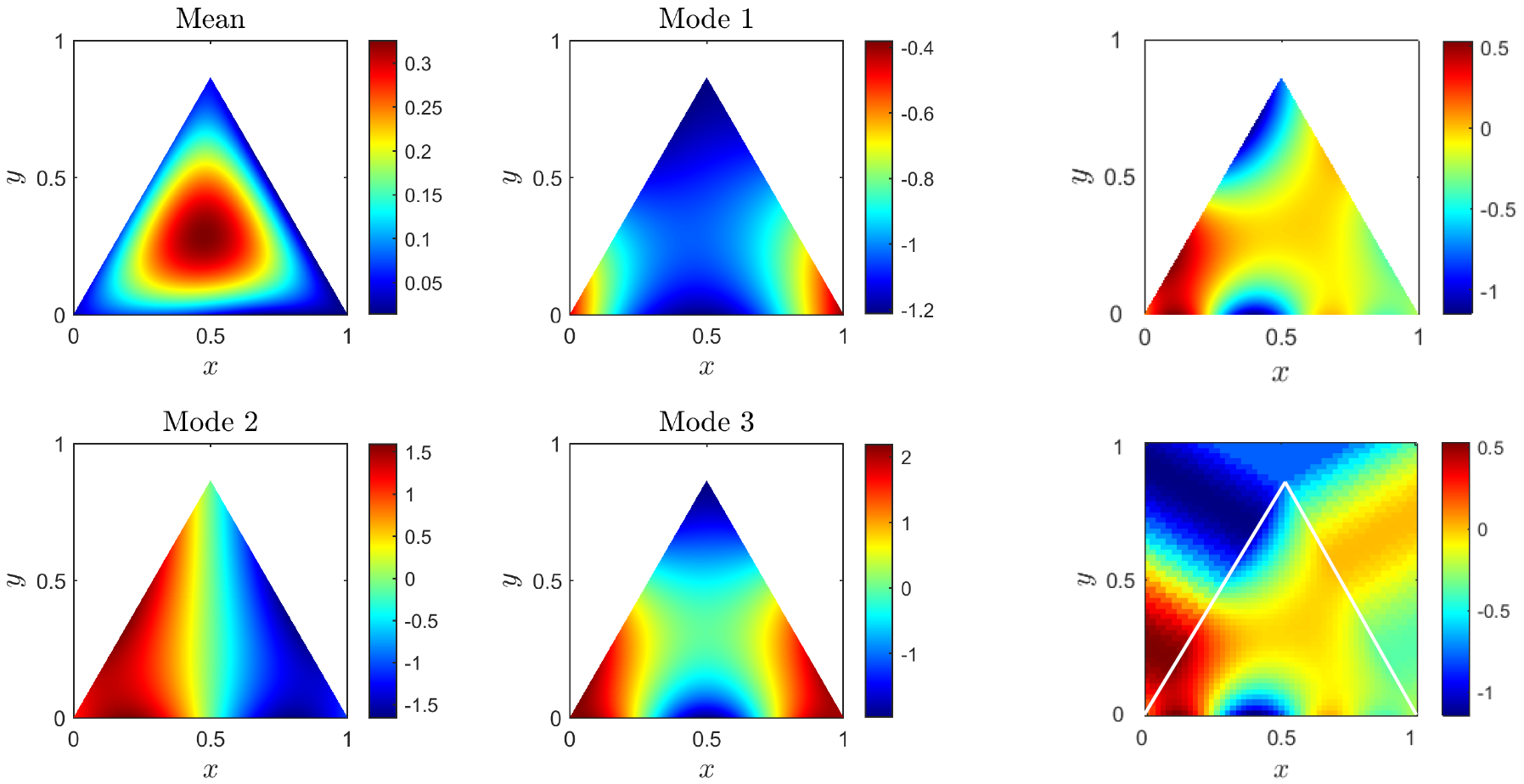}}
\caption{\textbf{Darcy flow in the domains of a pentagram and a triangle.}
(a) Darcy flow in a pentagram with a hole: two representative cases. For each case, Left: boundary condition; Right: pressure field. 
(b) Darcy flow in a triangular domain: two representative cases. Left: boundary condition; Right: pressure field. 
(c) Darcy flow in a triangular domain: Left: The first four POD modes used in POD-DeepONet. Right: An example of the augmented data for FNO training, where the padding is set by ``nearest neighbor''.
}
\label{fig:darcy_pentagram}
\end{figure}

\paragraph{Problem setup}
Here we consider the Darcy flow in a pentagram with a hole, where $K(x, y) = 0.1$ and $f = -1$. The following Gaussian process is employed to generate boundary conditions for each boundary in the pentagram:
\begin{equation}\label{eq:Gaussian_boundary}
\begin{split}
    h(x) &\sim \mathcal{GP}(0, \mathcal{K}(x, x')), \\
    \mathcal{K}(x, x') &= \exp[-\frac{(x - x')^2}{2 l^2}], ~l = 0.2,\\
    x, x' &\in [0, 1],
\end{split}
\end{equation}
Specifically, we draw one sample function from the above Gaussian process at each time, and compute the boundary values based on $x$ for each boundary of the pentagram. While the boundary values at the boundary of the hole is fixed as 1, i.e., $h(x) = 1$, for all test cases. In the present case, we generate 2000 sample functions from Eq.~\eqref{eq:Gaussian_boundary} as boundary conditions, then we employ the operator networks to learn the mapping from the boundary condition to the pressure field in the entire domain"
\[\cG: h(x,y)|_{\partial\Omega} \mapsto h (x,y). \] Two representative solutions with corresponding boundary conditions are displayed in Fig.~\ref{fig:darcy_pentagram}(a). 

\paragraph{Results}
As shown in Table \ref{tab:Darcy_rect}, the normalization of inputs/outputs helps to reduce the generation error in DeepONet, and hence, we utilize the normalization in DeepONet for all the Darcy problems in what follows. 

The results from the DeepONet, POD-DeepONet, and dgFNO+ are presented in Table \ref{tab:Darcy_geometry}. We note that FNO cannot be used for these problems, and instead we use dgFNO+, which is the combination of dFNO+ in Section~\ref{sec:FNO_domains} and gFNO+ in Section~\ref{sec:gfno}. In particular, the following grid size, i.e., $44 \times 44$, is employed in dgFNO+, which has almost the same nodes as in the original mesh, i.e., 1,938. As shown, the results from DeepONet and POD-DeepONet are quite similar, while both are about $2\%$ more accurate than the dgFNO+.

We further test an additional case using dgFNO+, in which we use a uniform grid with a resolution $50 \times 50$. The $L^2$ relative error for this case is $2.78 \pm 0.01\%$, which is better than the dgFNO+ with the resolution $44 \times 44$ but is still less accurate than the DeepONet and POD-DeepONet.  

\subsubsection{Darcy problem in a triangular domain}
\label{sec:darcy_triangular}

\paragraph{Problem setup}
The setup for the case considered here is the similar as in Section~\ref{sec:darcy_pentagram}, i.e., $K(x, y) = 0.1$, and $f = -1$. We also utilize the Gaussian process in Eq.~\eqref{eq:Gaussian_boundary} to generate the boundary conditions for each boundary of the triangular domain, and then use the operator networks to map the boundary conditions to the pressure field in the entire domain. Specifically, 861 nodes are employed in the numerical solver. Similarly, we also display two representative solutions with corresponding boundary conditions in Fig.~\ref{fig:darcy_pentagram}(b). 

\paragraph{Results}
As demonstrated in Table \ref{tab:Darcy_geometry}, both the DeepONet and POD-DeepONet are more accurate than the dgFNO+. POD-DeepONet achieves the best accuracy (0.18\%) among the three test approaches. The first four POD modes are demonstrated in Fig.~\ref{fig:darcy_pentagram}(c)(Left), where we note that each mode is scaled by its $L^2$-norm. For dgFNO+, we use a uniform grid with a resolution of $51\times51$ and the ``nearest neighbor'' in Section~\ref{sec:gfno}, and an example of the solution is illustrated in Fig.~\ref{fig:darcy_pentagram}(c) (Right).

\subsubsection{Darcy problem in a triangular domain with notch}
\label{sec:darcy_notch}

\begin{figure}[htbp]
\centering
\includegraphics[width=0.9\textwidth]{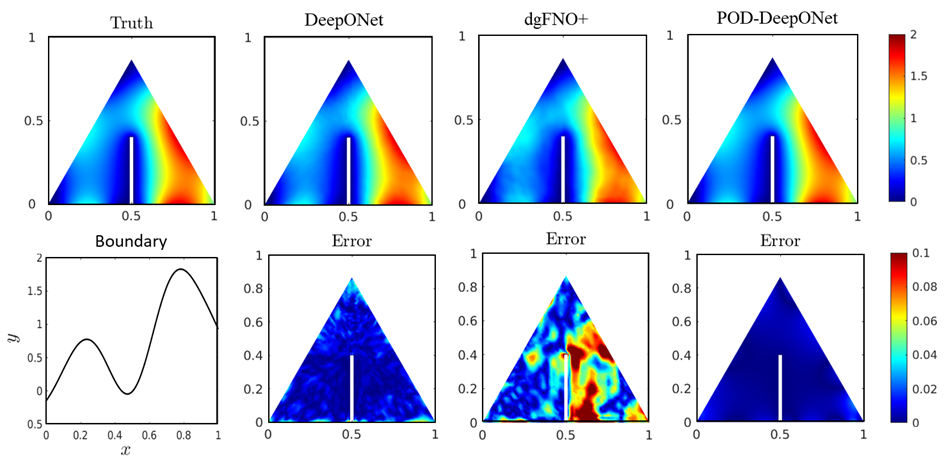}
\caption{\textbf{Darcy flow in a triangular domain with a notch.} For a representative boundary condition, the pressure field is obtained using DeepONet, dgFNO+, and POD-DeepONet. The prediction errors for the three operator networks are shown against the respective plots. The ground truth is simulated using the {\emph{PDE Toolbox}} in Matlab. The predicted solutions and the ground truth share the same colorbar, while the errors corresponding to each of the neural operators are plotted on the same colorbar.}
\label{fig:darcy_triagular_notch}
\end{figure}

\paragraph{Problem setup}
On the triangular domain discussed in Section~\ref{sec:darcy_triangular}, we now add a notch and also learn the operator mapping from the boundary conditions to the pressure field. The boundary conditions are generated using the Gaussian process discussed in Eq.~\eqref{eq:Gaussian_boundary}. A representative case is shown in Fig.~\ref{fig:darcy_triagular_notch}, where the predicted results and the errors corresponding to DeepONet, dgFNO+, and POD-DeepONet are shown for a specific boundary.

\paragraph{Results}
The $L_2$ relative error for this example is given in Table~\ref{tab:Darcy_geometry}. 
Amongst all the methods, POD-DeepONet outperforms the others, followed by DeepONet. The $L_2$ relative error for predicting the flow in a triangular domain with a notch is $2.64\pm 0.02\%$ using DeepONet, and is $7.82\pm0.03\%$ using dgFNO+. The lowest prediction error is $1.0\pm0.00\%$, obtained using POD-DeepONet.

\begin{table}[htbp]
    \centering
    \begin{tabular}{c|ccc}
    \toprule
     & \cref{sec:darcy_pentagram}  & \cref{sec:darcy_triangular} & \cref{sec:darcy_notch} \\
      & Darcy (Pentagram)    & Darcy (Triangular) & Darcy (Notch) \\
    \midrule
    DeepONet & 1.19$\pm$0.12\% & 0.43$\pm$0.02\% & 2.64$\pm$0.02\% \\
    FNO & --- & --- & --- \\
    \hdashline
    POD-DeepONet & \textbf{0.82$\pm$0.05}\%  & {\bf 0.18$\pm$0.02\%} & \textbf{1.00$\pm$0.00\%} \\
    dgFNO+ & 3.34$\pm$0.01\%   &  1.00$\pm$0.03\% & 7.82$\pm$0.03\% \\
    \bottomrule
    \end{tabular}
    \caption{\textbf{$L^2$ relative error for the Darcy flows in complex geometries.} dgFNO+ is the combination of dFNO+ and gFNO+.}
    \label{tab:Darcy_geometry}
\end{table}

\subsection{Multiphysics electroconvection problem}
\label{sec:electroconvection}

\paragraph{Problem setup}
Following the setup in~\cite{cai2021deepm}, we consider a 2D electroconvection problem, which is a multiphysics phenomenon involving coupling of the flow field with the  electric field, the cation and anion concentration fields. The full governing equations, including the Stokes equations, the electric potential and the ion transport equations, can be written as follows:
\begin{equation}
\begin{aligned}
\frac{\partial\bm{u}}{\partial t}    &= -  \nabla p + \nabla^{2}\bm{u} + \bm{f_{e}}, \\
\nabla \cdot \bm{u} &= 0,  \\
-2\epsilon^{2}\nabla^{2}\phi &= \rho_{e}, \\
\frac{\partial c^{\pm}}{\partial t} &= - \nabla \cdot \left( c^{\pm}\bm{u}-\nabla  c^{\pm} \mp c^{\pm} \nabla \phi \right), 
\end{aligned}
\label{eq:electroConvection_2D}
\end{equation}
where $\bm{u}$ and $p$ are the velocity and the pressure fields, respectively, and $\phi$ is the electric potential. Moreover, $c^{+}$ and $c^{-}$ are the cation and anion concentrations, respectively. Also, $\rho_{e}=(c^{+}-c^{-})$ is the free charge density, $f_{e}=-0.5\rho_{e}\nabla\phi/2\epsilon^{2}$ is the electrostatic body force, where $\epsilon$ is the Debye length. The investigated domain is defined as $\Omega: [-1,1]\times[0,1]$ with a regular mesh containing $101\times51$ grid points. 
By defining $\epsilon=0.01$, the electroconvection described in Eq.~(\ref{eq:electroConvection_2D}) becomes a steady flow, where the flow pattern is uniquely dependent on the electric potential difference ($\Delta \Phi$) acted on the upper and lower boundaries. 

\begin{figure}[htbp]
\centering
\includegraphics[width=\textwidth]{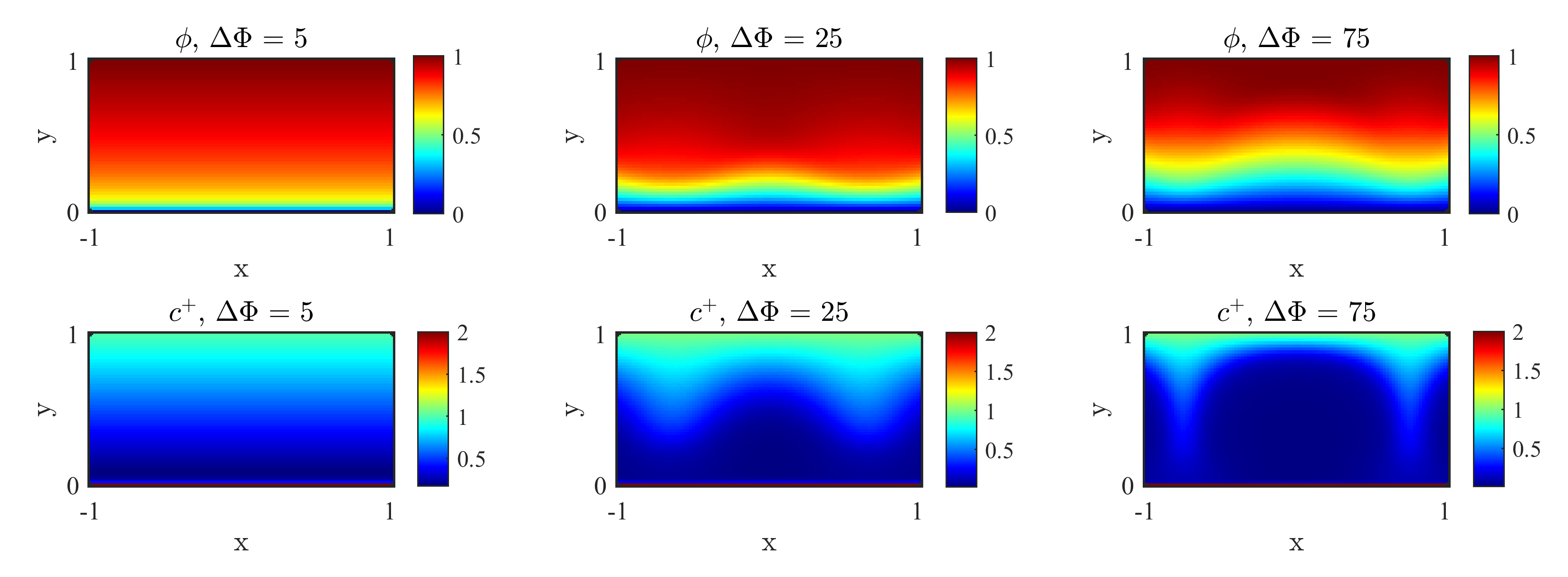}
\caption{\textbf{Electroconvection problem: three representative cases.} First row: electric potential field. Second row: cation concentration field. Note that there is a stiff boundary at $y=0$ in the concentration field, where the concentration drops from $c^{+}=2$ to approximately zero.}
\label{fig:electroconvection}
\end{figure}

In this problem, the operator networks are expected to learn the mapping from the 2D electric potential  field $\phi(x,y)$ to the 2D cation concentration field $c^{+}(x,y)$:
$$\cG: \phi(x,y) \mapsto c^{+}(x,y).$$ 
Different flow fields are generated by modifying the boundary condition of $\phi$, namely using $\Delta \Phi=5,10,\dots, 75$, which results in 15 steady states for network training. 
We also consider two unseen conditions, namely $\Delta \Phi=13.4$ and $\Delta \Phi=62.15$, which are applied for testing. The equations are solved by using a  high-order  spectral  element method. Three training cases of the electroconvection flow are demonstrated in Fig.~\ref{fig:electroconvection} and more details are included in~\cite{cai2021deepm}.

\paragraph{Results}
The evaluation errors for this example are given in Table~\ref{tab:electroconvection}, where we find that the testing errors of the different methods are all very small. Nevertheless, POD-DeepONet outperforms others, followed by DeepONet with data normalization. The reason that FNO performs relatively worse in this case is because of a stiff boundary at $y=0$ in the output field, where the concentration drops from $c^{+}=2$ to approximately zero. 
Note that the output resolution of this data set is $101\times51$, which means that there are 5151 sampling points for each output function used for training. However, it is reported in~\cite{cai2021deepm} that the DeepONet can be trained with much less data measurements (i.e., 800 random points for each output function) to achieve a similar accuracy of 0.49$\pm$0.04\%. This also demonstrates the flexibility of DeepONet training, since FNO requires the output representation to be on a regular mesh, while DeepONet does not have such a stringent requirement. 

\begin{table}[htbp]
\centering
\begin{tabular}{c|c}
\toprule
 & \cref{sec:electroconvection} \\
 & Electroconvection \\
\midrule
DeepONet w/o normalization & 0.26$\pm$0.04\% \\
DeepONet w/ normalization & 0.28$\pm$0.02\% \\
FNO w/o normalization & 1.00$\pm$0.01\% \\
FNO w/ normalization & 0.43$\pm$0.01\% \\
\hdashline
POD-DeepONet & \textbf{0.14$\pm$0.03}\% \\
\bottomrule
\end{tabular}
\caption{\textbf{$L^2$ relative error for the multiphysics electroconvection problem.}}
\label{tab:electroconvection}
\end{table}

\subsection{Wave propagation for continuous and discontinuous problems}
\label{sec:oscillatory}

\subsubsection{Advection equation}
\label{sec:advection}

\paragraph{Problem setup}

We consider the wave advection equation
$$\frac{\partial u}{\partial t} + \frac{\partial u}{\partial x}=0, \qquad x \in [0, 1], \quad t \in [0, 1],$$
with periodic boundary condition. We choose the initial condition as a square wave centered at $x=c$ of width $w$ and height $h$
\begin{equation*}
    u_0(x) = h1_{\{c-\frac{w}{2}, c+\frac{w}{2}\}},
\end{equation*}
where $(c, w, h)$ are randomly chosen from $[0.3, 0.7]\times[0.3,0.6]\times[1, 2]$. We first learn the mapping from the initial condition $u_0(x)$ to the solution at $t=0.5$:
$$\text{Case I:} \qquad \mathcal{G}_1: u_0(x) \mapsto u(x, t=0.5),$$
and then learn the operator mapping from the initial condition $u_0(x)$ to the solution $u(x, t) = u_0(x-t)$ of the whole domain
$$\text{Case II:} \qquad \mathcal{G}_2: u_0(x) \mapsto u(x, t).$$
The mesh of the solution is chosen as 40$\times$40 in space-time.
We also test a more complicated initial condition (Case III):
\begin{equation*}
    u_0(x) = h_1 1_{\{c_1-\frac{w}{2}, c_1+\frac{w}{2}\}} + \sqrt{\max(h_2^2-a^2(x-c_2)^2, 0)}.
\end{equation*}

\paragraph{Results}

\begin{figure}[htbp]
\centering
\includegraphics[width=\textwidth]{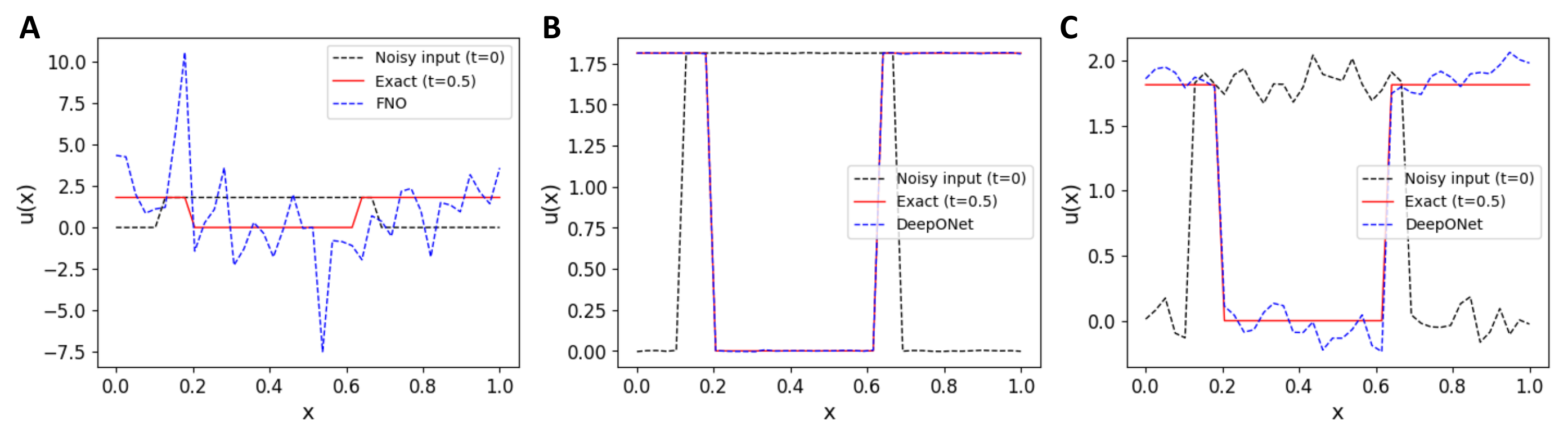}
\caption{\textbf{Advection equation (1D): one testing example of FNO and DeepONet for noisy inputs.} (\textbf{A}) FNO solution for an input with 0.1\% Gaussian noise. (\textbf{B}) DeepONet solution for input with 0.1\% Gaussian noise. 
(\textbf{C}) DeepONet solution for input with 5.0\% Gaussian noise.
The $L^2$ relative errors compared to the exact solution of clean input for (A--C) are 270\%, 0.36\% and  9.28\%, respectively. The functions in the plots are discretized with 40 uniform grid points.}
\label{fig:adv_1D_noise}
\end{figure}

For case I, DeepONet, POD-DeepONet and FNO all have errors $<1\%$, while POD-DeepONet achieves the smallest error ($<0.1\%$, Table~\ref{tab:oscillatory}). To evaluate the robustness of the operator networks, we add some small Gaussian noise to the input function during the testing stage. The noise level is defined as the percentage of the maximum value for the input function. Several examples are demonstrated in Fig.~\ref{fig:adv_1D_noise}, where we find that FNO is extremely sensitive to noise, i.e., it cannot handle 0.1\% Gaussian noise of the input. On the contrary, DeepONet is more robust, and it predicts perfectly for the input with 0.1\% noise. Moreover, it can also handle larger noise levels (as shown in Fig.~\ref{fig:adv_1D_noise}C).

For Cases II and III, DeepONet and POD-DeepONet still obtain good accuracy of the order of 0.1\% (Table~\ref{tab:oscillatory}). Because FNO cannot directly map from the 1D initial condition to the 2D solution, we considered the following three approaches.
\begin{itemize}
    \item \textbf{Brute-force FNO (2D):} We first use FNO in a brute-force way. Because the input $u_0$ is only a function of $x$, we directly repeat the same $u_0$ multiple times to match the 2D size of the output function, so that it is a valid 2D input. We find that the training loss is always very large $\sim\mathcal{O}(0.1)$ and cannot be improved, even if we do not truncate Fourier modes, use a large hidden dimension, and add more hidden layers. The $L^2$ relative error is $\sim$50\% (Table~\ref{tab:oscillatory}).
    \item \textbf{dFNO+1:} We use dFNO+ in  Section~\ref{sec:FNO_domains} for Case I using Method 1. Compared to the brute-force FNO above, here we also add the $t$ coordinate as input, and the optimization of FNO improves significantly. dFNO+1 achieves an error $< 1\%$ (Table~\ref{tab:oscillatory}), but it is still worse than DeepONet and POD-DeepONet.
    \item \textbf{dFNO+2:} We also use dFNO+ in Section~\ref{sec:FNO_domains} for Case I using Method 2, i.e., instead of using FNO in 2D, we test FNO in 1D with RNN for time-marching. We find that dFNO+2 is very hard to train, and the final result is unstable, as also observed in~\cite{li2020fourier}. In our 10 independent experiments, dFNO+2 gets stuck at a bad local minimum with ($L^2$ relative error $>$ 20\%) for 4 times. Here we truncate the FFT to the first 16 modes, but using more modes does not improve the accuracy or reduce the probability of training failure. We also note that the training cost of dFNO+2 is much more expensive than FNO in 2D.
\end{itemize}

\subsubsection{Linear instability waves in high-speed boundary layers}
\label{sec:instability_wave}

\begin{figure}[htbp]
    \centering
    \includegraphics{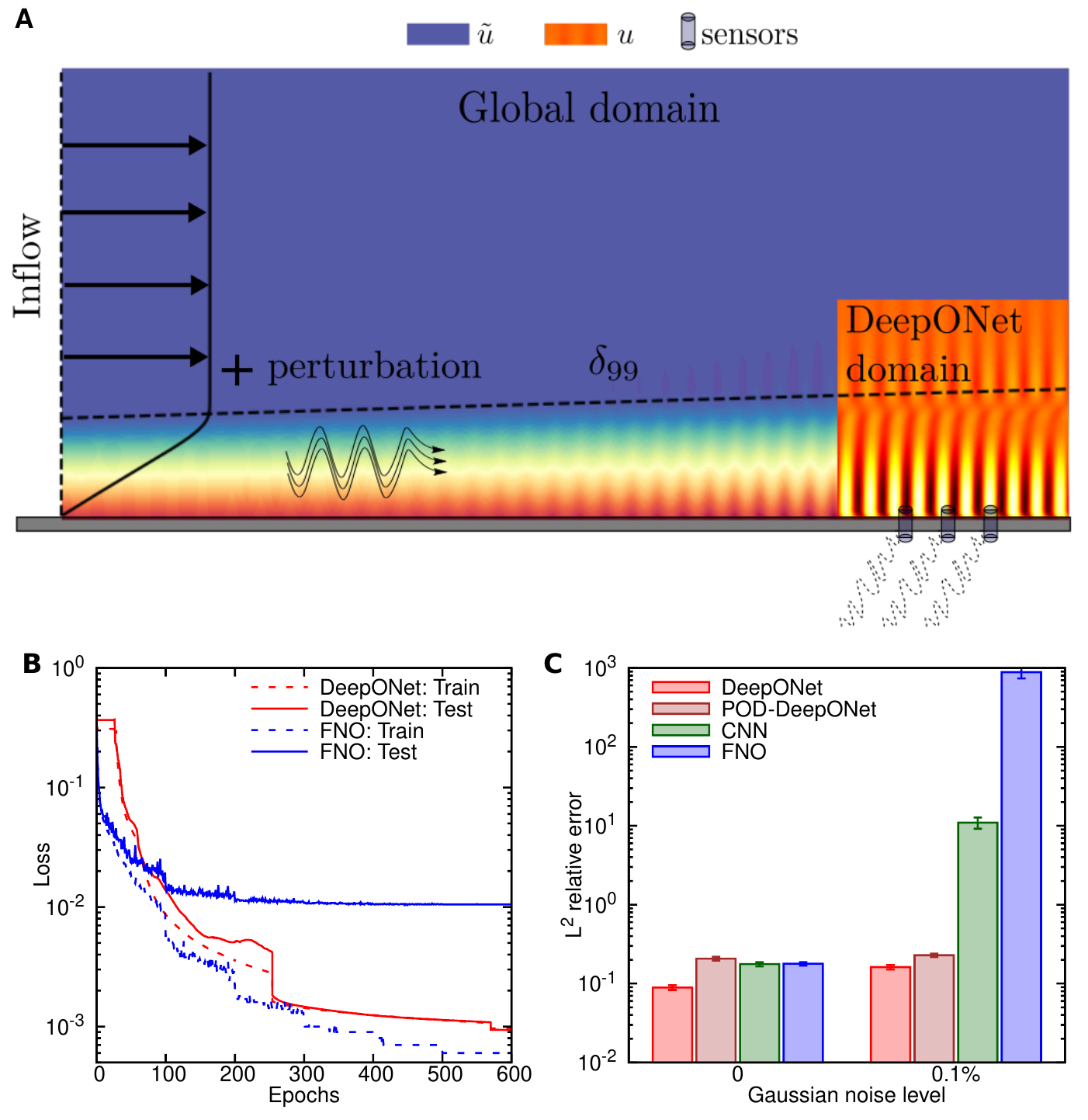}
    \caption{\textbf{Linear instability waves in high-speed boundary layers.} (\textbf{A}) Visualization of an instability wave in a spatially developing boundary layer. At the inlet to the computational domain, the base flow is superposed with instability waves.  The dashed line marks the 99\% thickness of the boundary layer. The objective is to accurately predict the downstream evolution of the instability wave. (\textbf{B}) The training and testing losses during the training process. (\textbf{C}) The $L^2$ relative errors of different networks for noiseless inputs and inputs with a 0.1\% Gaussian noise. Panel A is adapted from Ref.~\cite{di2021deeponet}.}
    \label{fig:instability_wave}
\end{figure}

\paragraph{Problem setup}
The early stages of transition to turbulence in high-speed aerodynamics often involve exponential amplification of linear instability waves. A visualization of an instability wave in a high-speed and spatially developing boundary layer is shown in Fig.~\ref{fig:instability_wave}A. We aim to predict the evolution of linear instability waves in a compressible boundary layer, i.e., how the upstream instability wave will amplify or decay within a region of interest downstream. Here, we consider small-amplitude instability waves, which can be accurately described by the linear parabolized stability equations (PSE) derived from the Navier-Stokes equations by decomposing the flow into the sum of a base flow and a perturbation.

The instability waves depends on the flow parameters. Here we consider air with Prandtl number $Pr=0.72$ and ratio of specific heats $\gamma = 1.4$. The free-stream Mach number is $Ma = 4.5$ and the free-stream temperature is $T_0 = 65.15K$. We set the inflow location of our configuration slightly upstream at $\sqrt{Re_{x_0}} = 1800$. When the perturbation frequency is $\omega$, the instability frequency is $\omega 10^6 / \sqrt{Re_{x_0}}$. In our dataset, the Reynolds number of the domain of integration spans from $1800^2$ to $2322^2$, and the problem was solved for 59 different perturbation frequencies in the range $[100, 125]$ with arbitrarily sampled phase. Moreover, the input function is defined on a mesh with resolution (20, 47), and the output function is on a mesh with resolution (111, 47). Because the output functions in the dataset differ in amplitudes by more than two orders of magnitude, a weighted MSE is used, where each loss term is weighted by the the amplitude of each function. More details about this problem and the dataset can be found in Ref.~\cite{di2021deeponet}.

\paragraph{Results}
This problem has been solved by DeepONet with 4 Fourier features in Ref.~\cite{di2021deeponet}. The accuracy of DeepONet is 8.90$\pm$0.60\%  (the forward case with two-mode combinations $F_2$ in Fig.~18 in Ref.~\cite{di2021deeponet}), and the training trajectory is shown in Fig.~\ref{fig:instability_wave}B (extracted from Fig.~15a in Ref.~\cite{di2021deeponet}). The output functions in the dataset differ by more than two orders of magnitude, and thus if we compute the PDE modes, the functions with large magnitude dominate. Hence, we first normalize all the functions such that the maximum value of each function is 1, and then compute the POD modes. This POD normalization makes POD-DeepONet work better, but POD-DeepONet still has a larger error (20.8$\pm$1.12\%) than DeepONet. A better strategy to compute the POD modes for problems with multiple scales need to be developed in the future. For FNO, because the input and outputs functions have different mesh resolution, we first use a linear interpolation to interpolate the input function from (20, 47) to (111, 47). FNO performs better than POD-DeepONet, but is much worse than DeepONet with Fourier features. Although DeepONet and FNO have a similar training error, DeepONet has a smaller testing error than FNO (Fig.~\ref{fig:instability_wave}B). In FNO, the generalization gap between the training and testing errors is more than one order of magnitude gap, while there is almost no generalization gap for DeepONet.

We further analyze the robustness of different networks to input uncertainties by adding a Gaussian noise of 0.1\% to the input functions during testing. We note that training data is still noiseless. We also add the result of CNN in Ref.~\cite{di2021deeponet} for the comparison. The errors of DeepONet and POD-DeepONet only increase slightly for noisy inputs and remain satisfactory, but the errors of CNN and FNO increase by two and four orders of magnitudes, respectively (Fig.~\ref{fig:instability_wave}C). This implies that the mapping that FNO has learned is unstable, hence future work should address this important issue.

\subsubsection{Compressible Euler equations with non-equilibrium chemistry}
\label{sec:euler}

\paragraph{Problem setup}
We consider one-dimensional inviscid high-speed flows with dissociation involving three species and non-equilibrium chemistry 
described by a general equation of state as studied in the works of~\cite{WangWeiShu2009JCP, ZhangShu2011JCP}:
\begin{equation*}
\left( 
\begin{array}{c}
\rho_1\\
\rho_2 \\
\rho_3 \\
\rho u\\
\rho E\\
\end{array}
\right)_t
+
\left( 
\begin{array}{c}
\rho_1 u\\
\rho_2 u\\
\rho_3 u\\
\rho u^2 + p\\
(E+p)u\\
\end{array}
\right)_x
=
\left( 
\begin{array}{c}
2M_1 \omega\\
-M_2 \omega\\
0\\
0\\
0\\
\end{array}
\right),
\end{equation*}
and 
\begin{equation*}
    \rho = \sum_{s= 1}^3\rho_s, \qquad p = RT\sum_{s=1}^3 \frac{\rho_s}{M_s}, \qquad E = \sum_{s=1}^3 \rho_s e_s(T) + \rho_1 h_1^0 +\frac{1}{2} \rho u^2,
\end{equation*}
where the enthalpy $h_1^0$ is a constant, $R$ is the universal gas constant, $M_s$ is the molar mass of species $s$, and the internal energy $e_s(T) = 3R/2M_s$ and $5R/2M_s$ for mono-atomic and diatomic species, respectively.
The rate of the chemical reaction is given by 
\begin{equation*}
\begin{aligned}
    \omega = 
    &\left( k_f(T)\frac{\rho_2}{M_2} - k_b(T)\left(\frac{\rho_1}{M_1}\right)^2
    \right) \sum_{s=1}^3 \frac{\rho_s}{M_s}, \quad
    k_f = CT^{-2}e^{-\mathcal{E}/T}, \\
    &k_b = k_f/e^{b_1+b_2\log z + b_3z + b_4 z^2 + b_5 z^3},
    \quad 
    z = 10000/T,
\end{aligned}
\end{equation*}
where $b_i,\, C$ and $\mathcal{E}$ are constants which can be found in \cite{gnoffo1989conservation, WangWeiShu2009JCP}.
The model involves three species, namely, $O_2, \, O$ and $N_2$ ($\rho_1 = \rho_O, \; \rho_2 = \rho_{O_2}$ and $\rho_3 = \rho_{N_2}$) with the reaction:
    $${O_2 + N_2 \Longleftrightarrow O +O +N_2}.$$
We consider the following initial condition:
\begin{equation*}
    T_0(x) = 8000K, ~~ 
    p_0(x) = \frac{p_L - p_R}{2}\left(1- \tanh(x/\eta)\right) + p_L,
\end{equation*}
where $\eta$ is a parameter in the range of $[0.02, 0.2]$, and $p_L$ and $p_R$ are constant left and right pressure levels. This is a multi-physics problem involving different quantities. Here, we only consider the pressure field and the Nitrogen density to demonstrate the comparison between different operator learners. The neural operators are expected to learn two mappings: from the initial condition $p_0(x)$ to the solution $p(x)$ at $t=0.0002$, and from $N_{2}(x,t=0)$ to  $N_{2}(x,t=0.0002)$. 

\paragraph{Results}

\begin{figure}[htbp]
\begin{center}
\includegraphics[width=\textwidth]{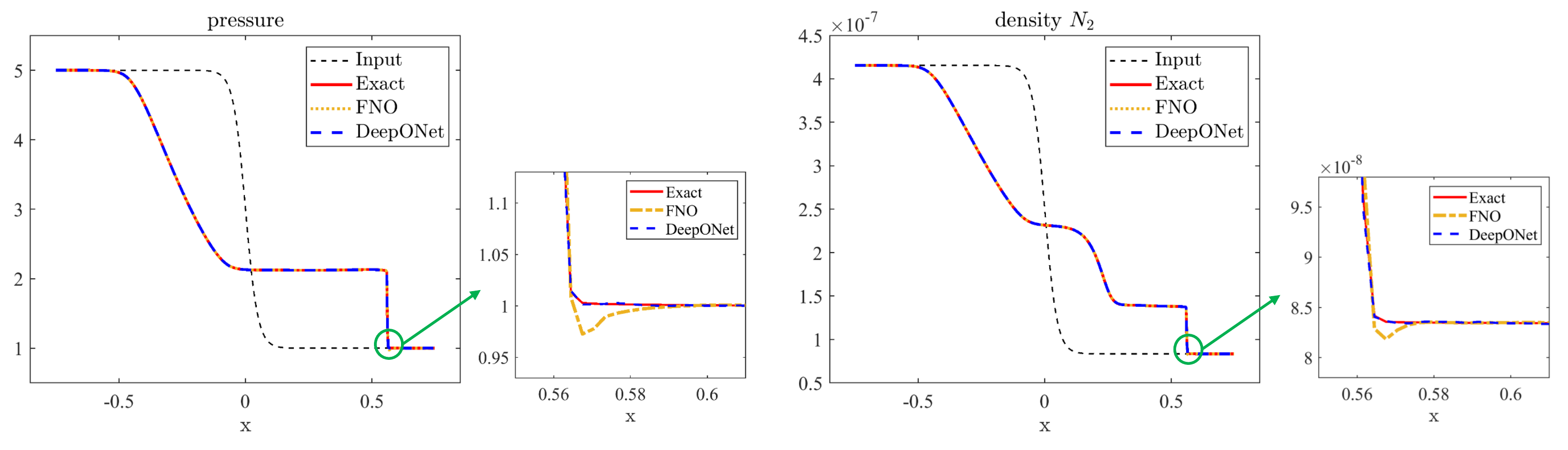}
\caption{\textbf{Compressible Euler equations: results of a testing example.} Left: neural operators for pressure. Right: neural operators for $N_2$ density. The black dash line is the initial condition. The red, orange, blue lines denote the exact solution, the FNO prediction and the DeepONet prediction, respectively.}
\label{fig:Euler}
\end{center}
\end{figure}

The $L^2$ relative errors of DeepONet and FNO are in Table~\ref{tab:oscillatory} and a testing example is demonstrated in Fig.~\ref{fig:Euler}. All the networks provide results with good accuracy, and POD-DeepONet performs the best. We note that the FNO relies on Fourier transformation, which is not very accurate for discontinuous functions. On the contrary, the DeepONet performs well for functions with discontinuity, as shown in the insets of Fig.~\ref{fig:Euler}. 

\begin{table}[htbp]
\small
\centering
\begin{tabular}{c|ccc}
\toprule
  & \cref{sec:advection} & \cref{sec:advection} &  \cref{sec:advection} \\ 
 & Advection (I) & Advection (II) & Advection (III) \\
\midrule
DeepONet   &  0.22$\pm$0.03\%  & 0.27$\pm$0.01\% &  \textbf{0.32$\pm$0.04}\%   \\
FNO    & 0.66$\pm$0.10\% & 54.4$\pm$0.00\% &  47.7$\pm$0.00\% \\
\hdashline
POD-DeepONet    & \textbf{0.04$\pm$0.00}\% & \textbf{0.08$\pm$0.00}\%  & 0.40$\pm$0.00\%  \\
dFNO+1 & --- & 0.22$\pm$0.01\% &  0.60$\pm$0.02\%  \\
dFNO+2 & --- & 3.89$\pm$1.26\% & 10.9$\pm$2.08\%  \\
\midrule 
 & \cref{sec:instability_wave} & \cref{sec:euler} & \cref{sec:euler} \\
 & Instability waves & Compressible Euler ($p$)  & Compressible Euler ($N_2$) \\
\midrule
DeepONet   & \textbf{8.90$\pm$0.60}\% & 0.068$\pm$0.011\%  & 0.043$\pm$0.006\% \\
FNO  & 17.8$\pm$0.92\%  & 0.076$\pm$0.005\% & 0.044$\pm$0.004\% \\
\hdashline
POD-DeepONet  & 20.8$\pm$1.12\%  & \textbf{0.020$\pm$0.004}\% &   \textbf{0.012$\pm$0.002}\%   \\
\bottomrule
\end{tabular}
\caption{\textbf{$L^2$ relative error of wave propagation for continuous and discontinuous problems in Section~\ref{sec:oscillatory}.}}
\label{tab:oscillatory}
\end{table}

\subsection{Predicting surface vorticity of a flapping airfoil}
\label{sec:flapping_airfoil}

\paragraph{Problem setup} 

\begin{figure}[htbp]
\begin{center}
\includegraphics[width=0.9\textwidth]{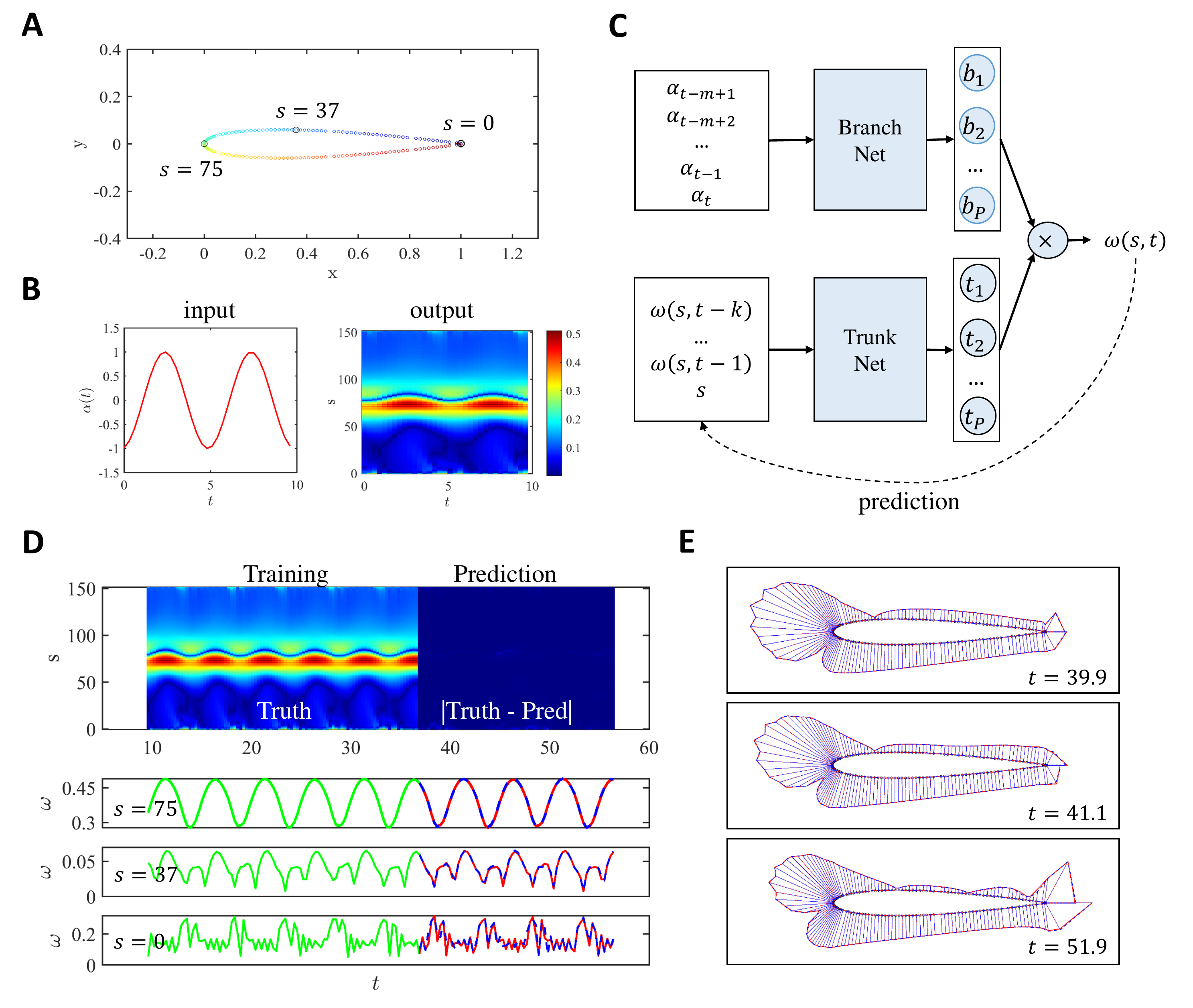}
\caption{\textbf{Predicting vorticity on the surface of a flapping airfoil.} (\textbf{A}) Geometry of NACA0012 airfoil.
(\textbf{B}) An example of the input function $\alpha(t)$ and output function $\omega(s,t)$ that are used to train the operator networks. (\textbf{C}) Schematic of the modified DeepONet, which includes a few historical states in the trunk net as the features to replace the time coordinate. 
(\textbf{D}) Testing result of the modified DeepONet. Top: 2D visualization in space-time; bottom: 1D signals at different surface locations. (\textbf{E}) Vorticity profiles on the airfoil surface at three time stamps. The red and blue colors in (D) and (E) represent the truth and the DeepONet prediction, respectively.
}
\label{fig:airfoil}
\end{center}
\end{figure}

Here, we predict the surface vorticity of a flapping airfoil based on the angle of attack. 
We perform simulation of the flow over a NACA0012 airfoil. The flapping airfoil setup is achieved by 
defining a oscillating inflow velocity, which is expressed as:
\begin{equation*}
\begin{aligned}
    u&=U_{\infty} \cos(\frac{\alpha_{0}\pi}{180} \times \frac{\sin(2f\pi t)+1.0}{2}),  \\
    v&=U_{\infty} \sin(\frac{\alpha_{0}\pi}{180} \times \frac{\sin(2f\pi t)+1.0}{2}), 
\end{aligned}
\end{equation*}
where $\alpha_{0}=15^{\circ}$ is the reference angle of attack (AOA), $f=0.2$ is the frequency and $U_{\infty}=1$. In this case, the Reynolds number based on the chord length is $Re=2500$. We can simply define the normalized time-dependent AOA as  $\alpha(t)=\sin(2f\pi t)$, which is used as the input of the learning algorithms hereafter. 
Specifically, we aim to map the AOA $\alpha(t)$ to the vorticity on the airfoil surface, which is denoted by $\omega(s,t)$ and $s$ is the location index on the surface. The airfoil geometry is illustrated in Fig.~\ref{fig:airfoil}A, where the surface of the airfoil is discretized by using 152 points. 

In this example, we only have one time-dependent signal, which covers about 56.4 time units. We apply approximately 36.6 time units for neural network training and the rest are used for prediction and validation. Moreover, we divide the training data into multiple independent signals, which involve two periods and have different phases. An example of the input-output functions for standard DeepONet and FNO training is illustrated in Fig.~\ref{fig:airfoil}B. By doing this, the training data contain 28 input-output pairs, while the testing data (about 20 time units) are composed of two signals whose phases are not included in training. 

For vanilla DeepONet and FNO, the operator can be expressed as: $\omega(s,t) =\cG(\alpha)(s,t)$. 
In this example, we also adopt a modified DeepONet with feature expansion, where a couple of historical states of the time signal are fed into the trunk net as the features and replace the time coordinate (Section~\ref{sec:feature}). Such an operator is denoted as:
$$\omega(s,t) =\cG(\alpha)(s,\omega(s,t-1),\dots,\omega(s,t-k)),$$
where $k$ denotes the number of historical states. We note that in the prediction stage, only the initial data of $\omega$ is given. The predictions from the network itself are concatenated and fed into the trunk net for the future predictions. A schematic of this DeepONet is shown in Fig.~\ref{fig:airfoil}C. 

\paragraph{Results} 

The $L^2$ relative errors of DeepONet and FNO for the testing data are given in Table~\ref{tab:airfoil}. Note that here we learn an operator mapping from a 1D function (i.e., function of time) to a 2D function (i.e., function of time and space). As mentioned above, FNO requires dimension augmentation for the input function in this case. If the input only involves the time coordinate, i.e., the brute-force way in Section~\ref{sec:advection}, then FNO fails to predict correctly the output (16.20\% error). However, when the input is augmented to involve both time and space coordinates (Section~\ref{sec:FNO_domains} Case I Method 1), the relative error of dFNO+ reduces to 3.56\%.

The vanilla DeepONet predicts the output with satisfactory result (3.65\%). In addition, when we apply the modified DeepONet with 5 historical states of the investigated quantity as the features, the error becomes even smaller (2.87\%). The prediction results of this modified DeepONet are illustrated in Figs.~\ref{fig:airfoil}D and E, where (D) shows the time-dependent signals and (E) shows the vorticity profiles over the airfoil surface. It is worth noting again that in the prediction process, the predictions from DeepONet are concatenated to the input of the trunk net in order to perform future 
evaluation. We can observe great consistency between the truth and the DeepONet prediction from the figures. 

\begin{table}[htbp]
\small
\centering
\begin{tabular}{c|c}
\toprule
 & \cref{sec:flapping_airfoil} \\
 & Flapping airfoil \\
\midrule
DeepONet & 3.65$\pm$0.02\% \\
FNO & 16.20$\pm$0.01\% \\
\hdashline
DeepONet (feature expansion) & \textbf{2.87$\pm$0.24}\% \\
dFNO+ & 3.56$\pm$0.10\% \\
\bottomrule
\end{tabular}
\caption{\textbf{$L^2$ relative error of predicting surface vorticity of a flapping airfoil.}}
\label{tab:airfoil}
\end{table}

\subsection{Navier-Stokes equation in the vorticity-velocity form}
\label{sec:ns}

\paragraph{Problem setup}
Following the problem setup in~\cite{li2020fourier}, we consider the 2D incompressible Navier-Stokes equation in the vorticity-velocity form:
\begin{equation*}
\begin{aligned}
    \partial_t \omega + \bm{u} \cdot \nabla \omega &= \nu \Delta \omega + \bm{f}, \quad &x\in[0,1]^2, ~ t\in[0,T], \\
    \nabla \cdot \bm{u} &= 0, \quad &x\in[0,1]^2, ~ t\in[0,T], \\
    \omega(x,0) &= \omega_{0}(x), \quad &x\in[0,1]^2,
\end{aligned}
\end{equation*}
where $\omega(x,y,t)$ and $\bm{u}(x,y,t)$ are the vorticity and velocity, respectively. The viscosity $\nu$ in our experiment is 0.001. The forcing term is defined as:
$$ f(x,y)= 0.1\sin(2\pi(x+y)) + 0.1\cos(2\pi(x+y)) $$
and the periodic boundary condition is imposed. 
The data set we use in this paper is identical to that in~\cite{li2020fourier}. We are interested in learning the operator mapping the vorticity field at the first ten time steps $t\in[0,10]$ to the vorticity at a target time step $T=20$. The solution is determined by modifying the initial condition $\omega_{0}$, which is given by a random field $\mu=\mathcal{N}(0, 7^{3/2}(-\Delta+49I)^{-2.5})$. The spatial resolution is fixed to be $64\times64$ both for training and testing. 

\paragraph{Results}
DeepONet in this example is imposed with periodic boundary condition in Section~\ref{sec:deeponet-bc} and a CNN is used to process the input function (namely $\omega(t_1-t_{10})$) in the branch net. For FNO, the input is the concatenation of $\omega(t_1-t_{10})$ and the grid points. The $L^2$ relative errors of DeepONet and FNO are illustrated in Table~\ref{tab:ns}, indicating that the accuracy of DeepONet and FNO are comparable in this case. First, we find that the networks without data normalization perform the worst (both giving more than 2\% errors). The data normalization improves the testing accuracy for both DeepONet and FNO, resulting in the errors of 1.78\% and 1.81\%, respectively. The best result is obtained by POD-DeepONet (1.71\%). 

\begin{table}[htbp]
\centering
\begin{tabular}{c|c}
\toprule
 & \cref{sec:ns} \\
 & Navier-Stokes \\
\midrule
DeepONet w/o normalization & 2.51$\pm$0.07\% \\
DeepONet w/ normalization & 1.78$\pm$0.02\% \\
FNO w/o normalization & 2.62$\pm$0.03\% \\
FNO w/ normalization & 1.81$\pm$0.02\% \\
\hdashline
POD-DeepONet & \textbf{1.71$\pm$0.03}\% \\
\bottomrule
\end{tabular}
\caption{\textbf{$L^2$ relative error for the Navier-Stokes equation in the vorticity-velocity form.}}
\label{tab:ns}
\end{table}

\subsection{Regularized cavity flows}
\label{sec:cavity}

Here we consider a two-dimensional lid-driven flow in a square cavity (i.e., $x, y \in [0, 1]$), which can be described by the incompressible Navier-Stokes equations as
\begin{align*}
    \nabla \cdot \bm{u} &= 0,\\ 
    \partial_t \bm{u} + \bm{u} \cdot \nabla \bm{u} &= -\nabla P + \nu \nabla^2 \bm{u}, 
\end{align*}
where $\bm{u} = (u, v)$ denotes the velocity in $x-$ and $y-$ directions, respectively; $P$ is the pressure; and $\nu$ is the kinematic viscosity. We consider two cases with different boundary conditions for the upper wall, i.e., a time-independent (Case A) and a time-dependent one (Case B). In particular, the boundary conditions are expressed as
\begin{align*}
    \mbox{Case A:}~ & u = U\left(1 - \frac{\cosh[r(x - \frac{L}{2})]}{\cosh(\frac{rL}{2})}\right), ~ v = 0,\\
    \mbox{Case B:} ~& u = U\left(1 - \frac{\cosh[r(x - \frac{L}{2})]}{\cosh(\frac{rL}{2})} +0.8 \sin(2\pi x) \sin(5t)\right) , ~ v = 0,
\end{align*}
where $U$, $r$ and $L$ are constant. Specifically, $r = 10$, $L = 1$ is the length of the cavity. In addition, the remaining walls are stationary in both cases. The aforementioned equations are then solved using the lattice Boltzmann method (LBM) \cite{meng2015multiple} to generate training data (see more details for the data generation in Section~\ref{sec:lbm}).

For Case A, we generate 100 velocity flow fields at different Reynolds numbers (Re = $U L / \nu$), i.e., spanning from 100 to 2080 with a step size 20. We then simulate the flow fields for 10 randomly generated Reynolds numbers within the range $[100, 2080]$, which are not seen in the training dataset and employ them as the testing dataset. We take the boundary condition on the upper wall as the input for the operator networks, and the corresponding output is the converged velocity field. 

\begin{figure}[H]
\centering
\subfigure[]{
\includegraphics[width=0.23\textwidth]{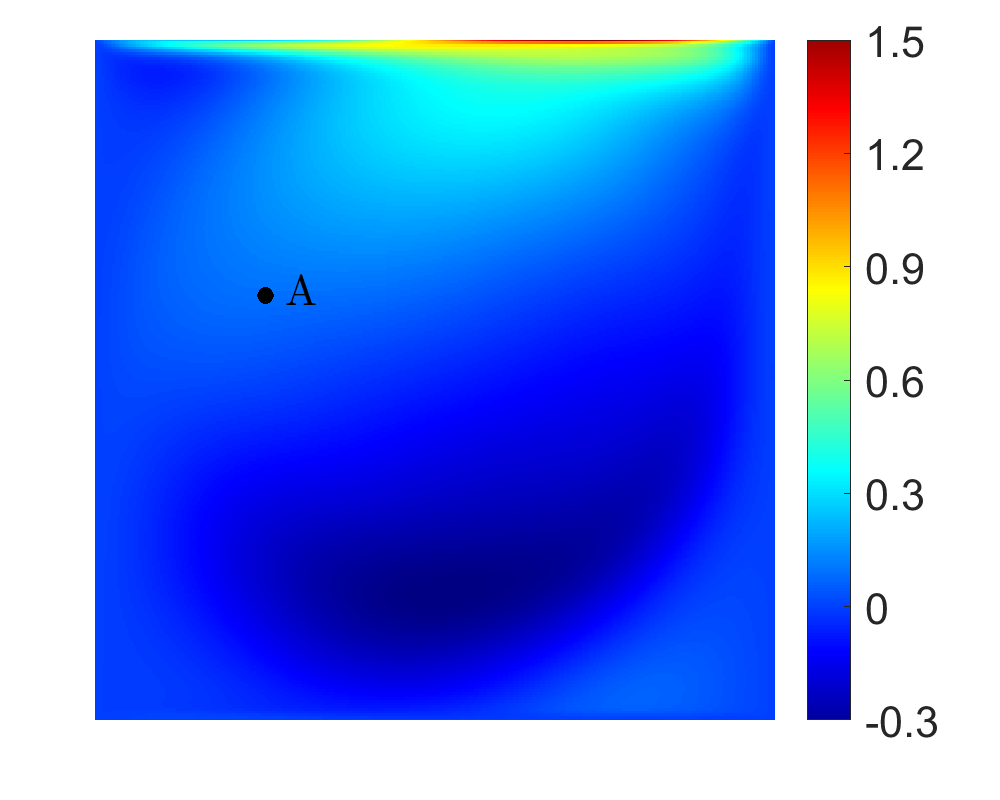}
\includegraphics[width=0.23\textwidth]{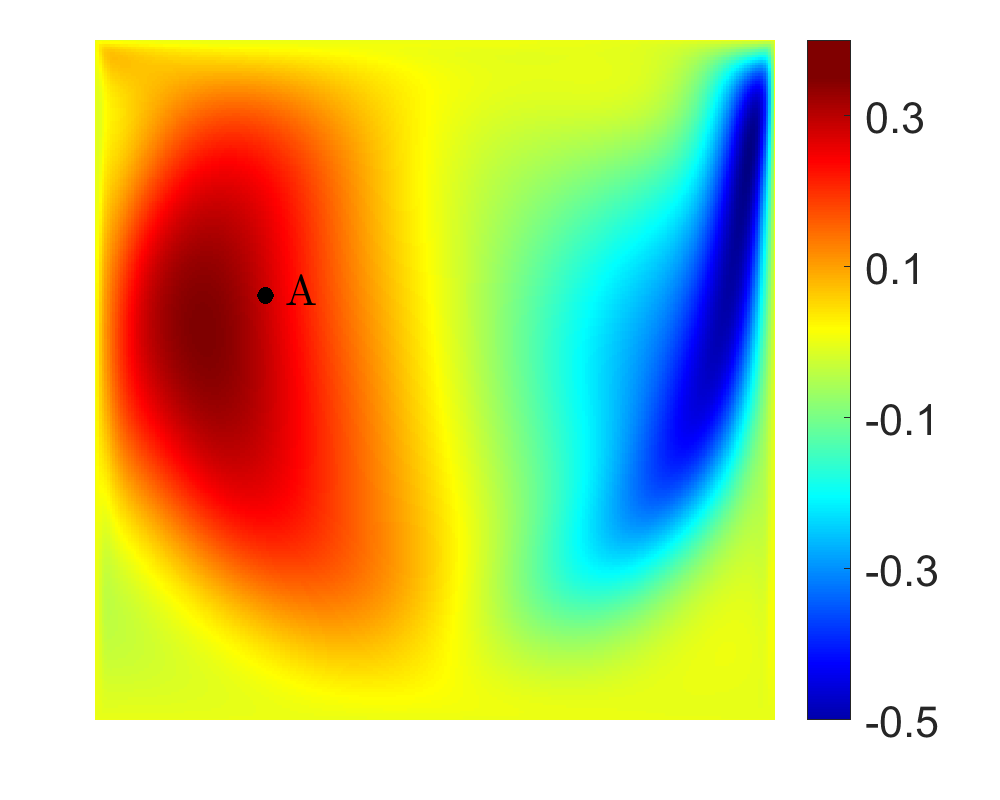}}
\subfigure[]{
\includegraphics[width=0.23\textwidth]{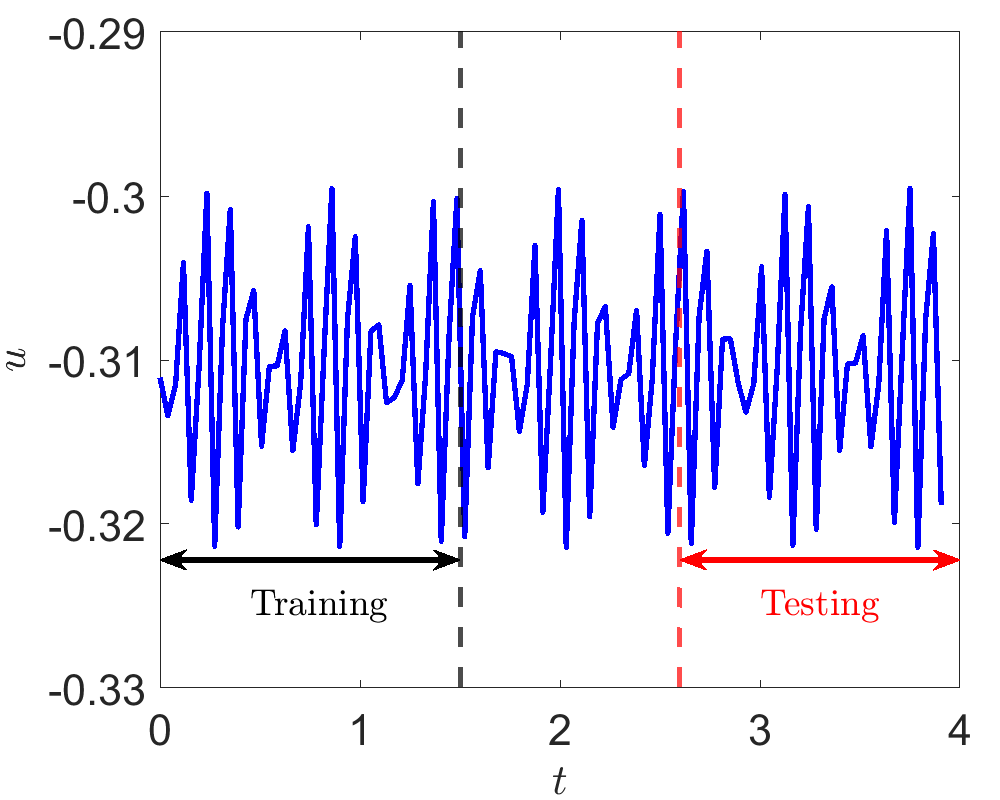}
\includegraphics[width=0.23\textwidth]{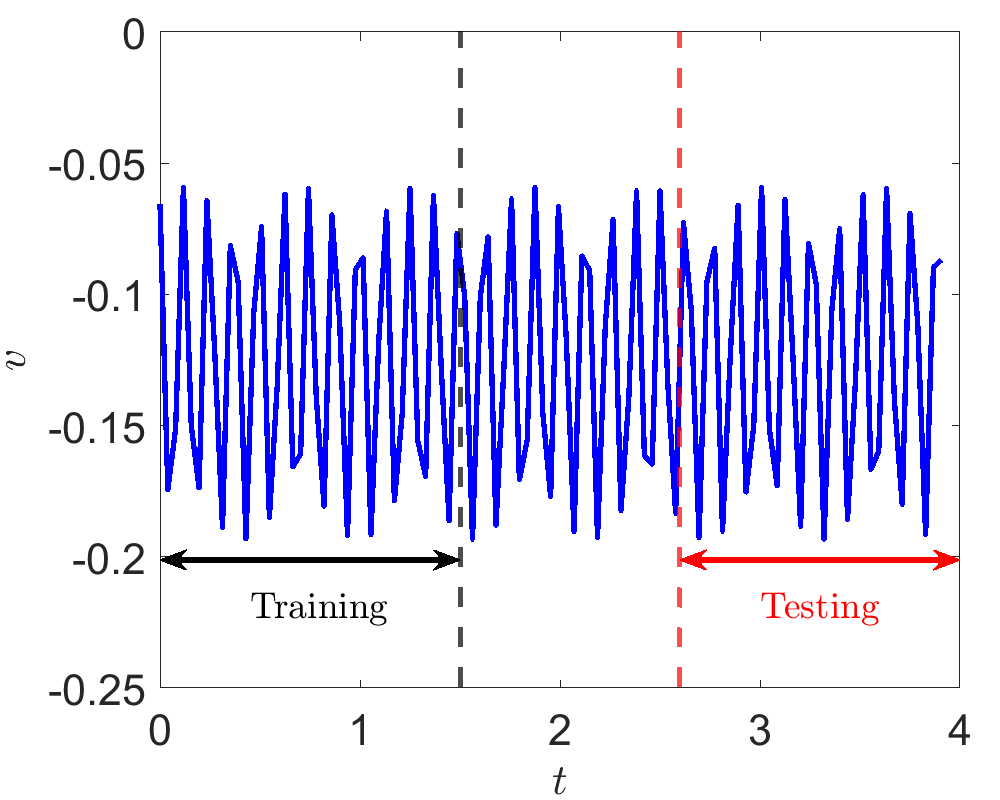}
}
\caption{\textbf{Unsteady cavity flows for Re $= 1000$ at $T = 1,500,000 dt$.}
(a) Velocity field obtained from LBM. Left: $u$, Right: $v$.
(b) Time series for the velocity at location $A$. 
$u$ and $v$ are normailized by $U$, and $t$ is normalized by $L/U$.
}
\label{fig:cavity}
\end{figure}

In Case B, we fix the Re  as 1000, and set the maximum iteration number in LBM as $1,500,000$,  i.e., $T = 1,500,000 dt$, where $dt$ is the time step used in LBM. We then save the velocity fields every 100 time steps in the last 10,000 iterations, suggesting that we have a total number of  100 snapshots for the velocity fields. Similarly, we train the operator networks to learn the mapping from the boundary condition to the corresponding velocity field. Specifically,  we consider the following two cases:
\begin{itemize}
    \item Unsteady I: We utilize the first 90 snapshots in the training and the last 10 as testing dataset.
    \item Unsteady II: We utilize the first 10 snapshots in the training and the last 10 as testing dataset. A representative velocity field for Re $= 1000$ is illustrated in Fig.~\ref{fig:cavity}.
\end{itemize}

Note that the boundary condition is assumed to be known in both Cases A and B. We then use $\sin(5t)$, which is introduced in the boundary condition in Case B as an extra feature for the networks to enhance the predicted accuracy. In particular, we employ $\sin(5t)$ as (1) an additional input for the Branch networks of DeepONet (by concatenating the boundary condition and $\sin(5t)$), and (2) an additional channel in dFNO+, i.e., we change the inputs from $(x, y, t, {u}_{bc})$ to $(x, y, t, \sin(5t), {u}_{bc})$, where $u_{bc}$ represents the velocity along the $x$ direction at the upper wall.  See more details on the feature expansion of DeepONet and FNO in Sections \ref{sec:feature} and \ref{sec:fno-feature}.

\paragraph{Results}


As displayed in Table~\ref{tab:cavity}: (1) For Case A, the results from dFNO+ are slightly more accurate than DeepONet, while POD-DeepONet achieves the best accuracy among them. (2) For Case B, DeepONet, dFNO+, and POD-DeepONet without feature expansion cannot provide accurate predictions for the velocity fields, and the relative errors from these three methods are quite similar. However, the extra feature can significantly reduce the predicted accuracy as compared to the results without feature expansion for all operator networks. In addition, DeepONet and dFNO+ with feature expansion have similar accuracy, and POD-DeepONet provides the most accurate predictions among all the methods. It is also interesting to observe that we can still get quite accurate predictions (around $2\%$ relative errors for all methods) for the velocity field as we reduce the number of training dataset from 90 to 10 with the feature expansion in Unsteady II. 

\begin{table}[htbp]
\centering
\begin{tabular}{c|ccc}
\toprule
 & \cref{sec:cavity} & \cref{sec:cavity} & \cref{sec:cavity} \\
 & Cavity (Steady) & Cavity (Unsteady I) & Cavity (Unsteady II)  \\
\midrule
DeepONet  & 1.20$\pm$0.23\%  & 14.4$\pm$0.53\% &  \\
FNO & --- & --- & --- \\
\hdashline
POD-DeepONet & \textbf{0.33$\pm$0.08}\% & 14.2$\pm$0.51\% &  \\
dFNO+ & 0.63$\pm$0.04\% & 15.0$\pm$0.17\% &  \\
DeepONet (feature expansion) & & 0.51$\pm$0.12\% & 2.24$\pm$0.21\% \\
POD-DeepONet (feature expansion)  & & \textbf{0.18$\pm$0.02\%} & \textbf{1.51$\pm$0.27\%} \\
dFNO+ (feature expansion) & & 0.56$\pm$0.03\% & 1.78$\pm$0.22\% \\
\bottomrule
\end{tabular}
\caption{\textbf{$L^2$ relative error for the regularized cavity flows.} Unsteady I: number of training data: 90; number of testing data: 10. Unsteady II: number of training data: 10; number of testing data: 10.}
\label{tab:cavity}
\end{table}






\section{Summary}
\label{sec:summary}

In this study, we have investigated the performance of two neural operators that have shown early promising results: the deep operator network (DeepONet) and the Fourier neural operator (FNO). The main difference between DeepONet and FNO is that DeepONet does not discretize the output, but FNO does. Moreover, DeepONet 
can employ any type of neural network architectures in the branch net whereas FNO has a fixed architecture, and hence DeepONet is more flexible than FNO in terms of problem settings and datasets (see comparison details in Table~\ref{tab:deeponet_fno}). Here, we have designed 16 benchmarks that have elements of industrial-complexity applications, e.g., unsteadiness, complex geometry, noisy data, and have generated data that will be available publicly so that interested researchers can test their own ideas against the results presented herein. In particular, we have shown that the vanilla DeepONet and the vanilla FNO may lead to suboptimal results in several of the 16 benchmarks that exhibit multiscale behavior and non-smooth solutions. To this end, we have proposed several extensions of both DeepONet and FNO (e.g., POD-DeepONet, dFNO+, gFNO+, and feature expansion) to either improve their accuracy or expand their capability to tackle diverse PDE based applications, especially for FNO to be able to deal with problems involving complex-geometry domains and mappings with different dimensionality of
the input-output spaces.

We have compared DeepONet and FNO both theoretically and computationally. We have theoretically showed that FNO and DeepONet of the same size exhibit the same accuracy when emulating the Cole-Hopf transformation for Burgers' equation. In particular, FNO in its continuous form can be thought of as a subcase of DeepONet with a specially-designed branch network and a discrete trigonometric basis to replace the trunk net. On the computational side, we have performed extensive experiments of 16 PDE problems, and we demonstrated that when proper extensions were employed, both DeepONet and FNO exhibited good accuracy for diverse applications, with similar performance in most problems. However, one major difference we found is the robustness to noise. In the examples of simulating the advection equation and instability waves, FNO was shown to be extremely sensitive to noise and totally failing to predict the solution even only when 0.1\% Gaussian noise was imposed on the testing data. On the contrary, DeepONet was much more robust and able to predict satisfactory results for relatively large noise levels.
This extreme sensitivity of FNO to noise implies that it may learn an unstable operator, which may not be able to generalize well even within the distribution of inputs.
In contrast, our experience even with the vanilla DeepONet is, e.g., see \cite{mao2021deepm}, that it generalizes well even for non-smooth solutions within and outside the distribution. For example, using the DeepONet as a building block in \cite{mao2021deepm} for modeling hypersonic flows, we were able to obtain very accurate results outside the Mach number training range $[8, 10]$, e.g., at Mach number 7 and 11, by assimilating only up to five new measurements of velocity and temperature. 

Training neural operators could be very expensive, but the theoretical works in \cite{lanthaler2021error,deng2021convergence} and \cite{kovachki2021universal} have shown that both 
DeepONet and FNO can break the curse of dimensionality in the input space for solution operators arising from the majority of partial differential equations. We have also demonstrated exponential convergence of the DeepONet error with respect to the size of the training data, see \cite{lu2021learning}, although for larger data sizes the convergence rate switches to algebraic due to the finite size of the neural network architecture. We believe that this is a possible direction for future work, i.e., to design new architectures that can sustain exponential convergence rate for all data sizes. Moreover, the demand for large data sets could be reduced significantly by incorporating physics-informed learning in the loss function of the neural operators as it was demonstrated in the works of \cite{wang2021learning} and \cite{goswami2021physics}. However, as demonstrated in
\cite{goswami2021physics}, a hybrid physics-data training is more effective for realistic problems of the type considered herein, and simply using the governing equations in the loss as suggested in \cite{wang2021learning} without using data may lead to erroneous results, e.g., for the compressible Euler equations considered in the current work.

\section*{Acknowledgements}

This work was supported by DARPA/CompMods HR00112090062, DOE PhILMs project (no.\ DE-SC0019453), and OSD/AFOSR MURI grant FA9550-20-1-0358.  Zhang was also partially supported by AFOSR under award number FA9550-20-1-0056.

\bibliography{main}

\clearpage
\appendix
\input{SI}

\end{document}

%% file: theory.tex
\section{Theoretical comparison}
\label{sec:theory}

In this section, we compare the universal approximation theorem of operators using DeepONet and FNO. We also compare their error estimates for the solution operator from the Burgers' equation. 

\subsection{Universal approximation theorem for operators}

\begin{thm}[\textbf{Generalized universal approximation theorem for operators}  \cite{lu2021learning}]
\label{thm:uat-cite-deeponets}
\label{thm:operator-univ-apprx-nn-uniform}
Suppose that $X$ is a Banach space, $K_1 \subset X$, $K_2 \subset \mathbb{R}^d$ are 
two compact sets in $X$ and $\mathbb{R}^d$, respectively, $V$ is a compact set in $C(K_1)$  
Assume that $\cG:\, V\to C(K_2) $   is a nonlinear continuous operator. 
Then for any $\epsilon>0$, there exist positive integers $m,p$, 
branch nets $b_k : V \to \mathbb{R}$,
and trunk nets  $t_k: \mathbb{R}^d\to\mathbb{R}$, and 
$x_1, x_2, \cdots, x_m \in K_1$, such that
\begin{equation*}
\sup_{v\in V} \sup_{x\in K_2}	\left|\cG(v)(x) -\sum_{k=1}^p \underbrace{b_k(v(x_1),v(x_2),\ldots,v(x_m))}_{branch}
	\underbrace{t_k(x)}_{trunk}
	\right|<\epsilon .
\end{equation*}
Furthermore, the functions $b_k$ and $t_k$ can be chosen as diverse classes of neural networks, satisfying the classical universal approximation theorem of functions, e.g., fully-connected neural networks, residual neural networks, and convolutional neural networks.
\end{thm}

This theorem was proved in \cite{chen1995universal} with two-layer neural networks. Also, the theorem 
holds when the Banach space $C(K_1)$ is replaced by $L^q(K_1)$ and $C(K_2)$ replaced by $L^r(K_2)$, $q,r\geq 1$.
Some extensions have been made in \cite{lanthaler2021error} for measurable operators, which contains discontinuous  operators that can be approximated by continuous operators.
The conclusion can be readily extended to the case of vector (multiple) output functions.

The FNO  for operator regression is proposed in 
\cite{li2020fourier}, and the corresponding universal approximation theorem is presented in   \cite{kovachki2021universal}. We present the theorem for completeness.
Define $\mathbb{T}^d= [0,2\pi]^d$ and the Hilbert space with smooth index $s$ by  $H^{s}\left(\mathbb{T}^{d} ; \mathbb{R}^{d_{v}}\right)$ 
for $\mathbb{R}^{d_{v}}$-valued functions defined on $\mathbb{T}^{d} $.
Let $\cV(\mathbb{T}^d;\Real^{d_v}) $ be a Banach space of $\Real^{d_v}$-valued functions defined on $\mathbb{T}^d$.
Let $P$ be a lifting operator from $\cV(\mathbb{T}^d;\Real^{d_v}) \to \cU(\mathbb{T}^d;\Real^{d_a})$ and the projection $Q: \cU(\mathbb{T}^d;\Real^{d_a}) \to \cU(\mathbb{T}^d;\Real^{d_u})$.
Let $\cI_N$ denote the Fourier interpolation operator, i.e., $\cI_N f (x)= \sum_{i=1}^N f(x_i) L_i(x)$, where $x_i$'s are the Fourier collocation points on $\mathbb{T}^d$ and $L_i(x)$ are corresponding Lagrange interpolation trigonometric polynomials.
Therefore the output of the $l$th Fourier block is defined as
\begin{equation*}
 \cL_l(z)(x) =
 \sigma\bigg(
 W_l z(x) + b_{l}(x) + \cF^{-1}\Big(R_l(k) \cdot \cF(z)(k)\Big)(x)\bigg).
\end{equation*}
Here, $W_l\in \Real^{d_l\times d_l}$ and 
the  function 
$b_{l}(x)$ is $\Real^{d_l}$-valued, and the coefficients $R_l(k) \in \Real^{d_l\times d_l}$ define a convolution operator via the Fourier transform.
Then, the FNO in its continuous form is defined as follows 
\begin{align} \label{eq:form-pfno}
\mathcal{F}(v) =Q \circ \cI_N \circ \cL_L \circ \cI_N \circ \dots \circ \cL_1 \circ \cI_N \circ P(v).
\end{align}


\begin{thm}[Universal approximation theorem of FNO \cite{kovachki2021universal}]
Let $s,s'\geq 0$ and $\Omega\subset \mathbb{T}^d$ be a domain with Lipschitz boundary. Let ${\cG}: H^{s}\left(\Omega; \mathbb{R}^{d_{v}}\right) \rightarrow$ $H^{s^{\prime}}\left(\Omega ; \mathbb{R}^{d_{u}}\right)$ be a continuous operator. Let $V \subset H^{s}\left(\Omega ; \mathbb{R}^{d_{v}}\right)$ be a compact subset. Then for any $\epsilon>0$, there exists a FNO of the  form \eqref{eq:form-pfno}, $\mathcal{F}: H^{s}\left(\mathbb{T}^{d} ; \mathbb{R}^{d_{v}}\right) \rightarrow H^{s^{\prime}}\left(\mathbb{T}^{d} ; \mathbb{R}^{d_{u}}\right)$ 
such that
$$
\sup _{v \in V}\|{\cG}(v)-\mathcal{F}(v)\|_{H^{s^{\prime}}} \leq \epsilon.
$$
\end{thm}
\noindent The conclusion above can be found in Theorem 2.15 of  \cite{kovachki2021universal}.

Both DeepONet and FNO suffer from \textbf{the curse of dimensionality} if one uses ReLU or tanh networks for Lipschitz continuous operators, due to the approximation capacity of these networks for high dimensional inputs $(v(x_1), v(x_2), \ldots,v(x_m))$. However, rates of convergence for the DeepONet are obtained for \cite{deng2021convergence,lanthaler2021error}, for some solution operators from PDEs.  Next, we compare error estimates of DeepONet and FNO for 1D Burgers equations with periodic boundary conditions. As will be shown, the inputs of the operator will be first approximated by some numerical methods such that one has a possibly high-dimensional inputs for the neural  network operators. This will introduce approximation errors in the inputs and thus in the outputs. Then extra  errors will be induced by the network approximation  emulating the analytical (after approximation of the input function) method of the solution. 

\subsection{1D Burgers' equation}

Consider the  Burgers' equation with  periodic boundary condition 
\begin{eqnarray}
\left\{\begin{array}{ll}
u_t + u  u_x = \kappa u_{xx}, \ \ (x,t) \in \Real \times(0,\infty),\ \ \kappa>0,\\ 
u(x-\pi,t) = u(x+\pi,t ), \quad
u(x,0)=u_0(x).
\end{array}\right.
\label{eqtburgers}
\end{eqnarray}
Let $M_0,\,M_1>0$. Define
\begin{eqnarray*}
 \mathcal{S} = \mathcal{S}(M_0, M_1) := \{ v\in W^{1,\infty}(-\pi,\pi): \norm{v}_{L^{\infty}} \leq M_0, \| \partial_x v\|_{L^{\infty}} \leq M_1, \bar{v}:=\int_{-\pi}^{\pi}v(s)\,ds = 0 \}.
\end{eqnarray*}
We consider the Fourier interpolation of $u_0$, 
$u(x,0)=u_0(x) = \sum_{j=0} ^{m-1} u_{0,j} L_j(x)$,
where  
$-\pi = x_0 < x_1<\cdots <x_m = \pi$, $u_{0,j} = u_0(x_j)$, and $L_j(x)$ is  the Lagrange interpolation trigonometric polynomials at $\mathbf{u}_{0,m} = (u_{0,0}, u_{0,1},\cdots, u_{0,{m-1}})^\top$.

 Define $x^l_j = x^0_j + 2\pi l$, $j=0,1,\cdots,m$, for each $l \in \mathbb{Z}$. Then $\{ x_j^l\}_{j=0}^m$ form  a partition of $[-\pi+2\pi l, \pi+2\pi l)$. For simplicity, we denote $x_j = x^0_j$. 
To make sure that $v_0(x)$ in \eqref{burgers-ch-pbc}  is $2\pi$-periodic, we require that the initial condition has zero mean in a period
$\bar{u}_0:=\int_{-\pi}^{\pi}u_0(s)ds = 0$.
Then, by the Cole-Hopf transformation, the solution to \eqref{eqtburgers} can be written as 
\begin{eqnarray}\label{burgers-ch-pbc}
u = \frac{-2\kappa v_x}{v}, \quad \hbox{where} \quad 
\left\{\begin{array}{ll}
v_t=\kappa v_{xx},\\
v(x,0) = v_0(x) = \exp\Big(-\frac{1}{2\kappa} \int_{-\pi}^x u_0(s) ds \Big).
\end{array}\right. 
\end{eqnarray}
Since $0 < \exp(-\frac{\pi \| u_0\|_{\infty}}{\kappa}) \leq v_0(x) \leq \exp(\frac{\pi \| u_0\|_{\infty}}{\kappa})$, the solution $u$ can be written explicitly as 
\begin{eqnarray}
u(\mathbf{x}) = \mathcal{G}(u_0)(\mathbf{x})&:=&-2\kappa \frac{ \int_{\mathbb{R}} \partial_x \mathcal{K}(x,y,t) v_0(y) dy}{ \int_{\mathbb{R}} \mathcal{K}(x,y,t) v_0(y) dy},\quad \mathbf{x} = (x,t),
\label{burgersexcslt}
\end{eqnarray}
where $\mathcal{K}(x,y,t) = \frac{1}{\sqrt{4\pi\kappa t}} \exp\Big(- \frac{(x-y)^2}{4\kappa t} \Big)$ is the heat kernel. 
It can be readily checked that $u(\mathbf{x})$ is a unique solution to \eqref{eqtburgers}.

We may obtain a neural network for operators of regression of $\cG$ by first approximating it with classical numerical methods and then emulating these methods using neural networks. %
Let us first approximate $\mathcal{G}$. Define $\mathbf{V}_m :=  \mathbf{V}(\mathbf{u}_{0,m})=(V_0, V_1,\cdots, V_{m-1})^\top$, where $V_0=1$ and $V_j =  \exp(-
\sum_{k}\int_{-\pi}^{x_j}L_k (y)\,dy \frac{u_0(y_k)} {2\kappa}  )$, $j=1,\cdots,m-1$. 
Define $\cG_m (\mathbf{u}_{0,m}) (\mathbf{x})= \Big( \widetilde{\cG}_m \circ \mathbf{V}  (\mathbf{u}_{0,m})\Big)(\mathbf{x})$ where 
\begin{eqnarray}
\cG_m  (\mathbf{u}_{0,m}) 
(\mathbf{x}) = \frac{-2\kappa \int_{\mathbb{R}} \partial_x \mathcal{K}(x,y,t) (\mathcal{I}_m v_0)(y) dy }{\int_{\mathbb{R}} \mathcal{K}(x,y,t)( \mathcal{I}_m v_0)(y) dy}
= \frac{V_0 c_0^1(\mathbf{x}) + V_1  c_1^1(\mathbf{x}) + \cdots + V_{m-1}c_{m-1}^1(\mathbf{x}) }{ V_0 c_0^2(\mathbf{x}) + V_1 c_1^2(\mathbf{x}) + \cdots + V_{m-1} c_{m-1}^2(\mathbf{x}) },\label{defrational}
\end{eqnarray}
where $\widetilde{\cG}_m$ is a rational function with respect to $\mathbf{V}_m$, with  both the numerator and the denominator being $m$-th degree $m$-variate polynomials with $m$ terms  and  $\mathcal{I}_m $ is  the Fourier interpolation operator and  for $j=0,\cdots, m-1$, 
\begin{eqnarray*}
	c_{j}^1(\mathbf{x}) = -2\kappa  \int_{0}^{x} \Big( \sum_{l\in \mathbb{Z}} \partial_x \mathcal{K}(x,y+2\pi l, t)  \Big) L_j (y) \, dy, \quad 	c_{j}^2(\mathbf{x}) = \int_{0}^{x} \Big( \sum_{l\in \mathbb{Z}} \mathcal{K}(x,y+2\pi l, t)  \Big) L_{j}(y)  \,dy.
\end{eqnarray*}
By the Lipschitz continuity of the solution operator 
$\cG(\cdot)$ of the Burgers equation, we have the following estimate.

\begin{thm}\label{thm:oprterror1d}
Let $u_0\in \mathcal{S}'=\set{u_0|_{[-\pi,\pi]}\in \mathcal{S}: u_0  \text{ grows at most quadratically at } \infty}$. Let $\mathcal{G}(u_0)(\mathbf{x})$ and $\cG_m  (\mathbf{u}_{0,m})( \mathbf{x})$ be defined in \eqref{burgersexcslt} and \eqref{defrational}, respectively.  Suppose $h$ is small enough. Then there is a uniform constant $C$ depending only on $t$, the lower and upper bounds $M_0$ and 
$M_1$ and $\kappa$, such that for any $t \in  (0,+\infty)$, we   have 
	\begin{eqnarray*}
\sup_{u_0\in \mathcal{S}'}	\norm{ \mathcal{G}(u_0)(\cdot,t) - \cG_m  (\mathbf{u}_{0,m})(
		\cdot,t)}_{L^2(-\pi,\pi)}
		 \leq C  h, \text{ where }  h= 2\pi/m.
	\end{eqnarray*}
\end{thm}
\begin{proof}
The proof can be found in Section~\ref{sec:proof}.
\end{proof}

Second, we emulate   $\cG_m$ by a neural network. To this end, it is important to realize that a rational function can be approximated by a ReLU network according to the following theorem.
\begin{thm}[\cite{telgarsky2017neural}]\label{thm:rational-apprx-ReLU}
	Let $\varepsilon \in (0,1]$ and nonnegative integer $k$ be given. Let $p:[0,1]^d\rightarrow [-1,1]$ and $q:[0,1]^d\rightarrow [2^{-k},1]$ be polynomials of degree $\leq r$, each with $\leq s$ monomials. Then there exists
	a ReLU network $f$ of size (number of total neurons)
$	\mathcal{O}\Big(k^7 \ln(\frac{1}{\varepsilon})^3 + \min\{ srk\ln(sr/\varepsilon), \ sdk^2\ln(dsr/\varepsilon)^2 \} \Big)
$  
	such that 
	$\sup_{x\in[0,1]^d} \Big| f(x) - \frac{p(x)}{q(x)}\Big| \leq \varepsilon.
	$
\end{thm}

Observing that $\cG_m (\mathbf{u}_{0,m})=\cG'(\mathbf{V}_m)$ is a rational function with respect to $\mathbf{V}_m$ while   $\mathbf{V}_m$ is an exponential function in   $\mathbf{u}_{0,m}$. Then by approximation of rational polynomials in Theorem \ref{thm:rational-apprx-ReLU}, there exists a ReLU network of size $\mathcal{O}(m^2\ln(m))$ ($s=r=m$ in Theorem \ref{thm:rational-apprx-ReLU}) to obtain accuracy of $\mathcal{O}(m^{-1})$ for each $x$. 
For the approximation of  `exp' by ReLU networks,  we only need  a network size of  $\mathcal{O}(\ln(m))$ to obtain accuracy of $\mathcal{O}(m^{-1})$ , according to \cite{OpsSZ2019}. 
Thus,  we need a ReLU network of size $\mathcal{O}(m^3\ln(m))$ to emulate $\cG_m(\mathbf{u}_{0,m})$ and denote this ReLU network by  
$\cG_m^{\mathbb{N}}(\mathbf{u}_{0,m})$.

The network $\cG_m^{\mathbb{N}}(\mathbf{u}_{0,m})(x_j)$  can be viewed as FNO, where for all the kernels $R_l \equiv 0$. Thus, the FNO of size $\mathcal{O}(m^3\ln(m))$ can achieve the accuracy of $\mathcal{O}(m^{-1})$. 
According to \cite{deng2021convergence}, the network $\cG_m^{\mathbb{N}}(\mathbf{u}_{0,m})(x_j)$ serves as a  branch net while only a ReLU network of size $\mathcal{O}(\ln m)$ is needed for the trunk net. The size of the DeepONet  is thus $\mathcal{O}(m^3\ln(m))$. In conclusion,  we find that 
\begin{itemize}
    \item both FNO and DeepONet of size $\mathcal{O}(m^3\ln(m))$ can achieve the accuracy of $\mathcal{O}(m^{-1})$.
\end{itemize}

There are many ways to approximate the solution operator and emulate the corresponding numerical methods. Thus, the estimation of sizes of networks may be not optimal\footnote{To make a fair comparison between DeepONet and FNO, we always emulate the same numerical methods.}. However, we can always connect FNO  $ \cF(v)$ and DeepONet by  
$
 \sum_{i=1}^N \cF(v) (x_j) L_j(x),
$  
 where $L_j(x)$ is a basis of interpolation type, i.e., 
 $L_j(x_j)=1$ and $L_j(x_i)=0$ for all $i\neq j$. Depending on the underlying problems, the basis $L_j(x)$ can be trigonometric polynomials, piecewise or global algebraic polynomials, splines, neural networks, etc. {\em In other words, FNO can be thought of as a special form of branch network in DeepONet.}

%% file: SI.tex
\renewcommand{\thesection}{S\arabic{section}}
\renewcommand{\thefigure}{S\arabic{figure}}
\renewcommand{\thetable}{S\arabic{table}}
\setcounter{figure}{0}
\setcounter{table}{0}

\section{Proof of Theorem~\ref{thm:oprterror1d}}
\label{sec:proof}

\begin{proof}
By the fact that  
$\partial_x\mathcal{K}(x,y,t)=-\partial_y  \mathcal{K}(x,y,t)$ and integration by parts, we have 
	\begin{eqnarray*}
		\mathcal{G}(u_0)(\mathbf{x}) - \cG_m  (\mathbf{u}_{0,m}; \mathbf{x}) &=& u(\mathbf{x})\frac{ \int_{\mathbb{R}}  \mathcal{K}(x,y,t) (\mathcal{I}_m v_0 - v_0)(y) dy}{\int_{\mathbb{R}} \mathcal{K}(x,y,t)( \mathcal{I}_m  v_0)(y) dy} - \frac{2\kappa \int_{\mathbb{R}}  \partial_x \mathcal{K}(x,y,t) (\mathcal{I}_m v_0 - v_0)(y) dy}{\int_{\mathbb{R}} \mathcal{K}(x,y,t)( \mathcal{I}_m  v_0)(y) dy}\\
		&=&  u(\mathbf{x})\frac{ \int_{\mathbb{R}}  \mathcal{K}(x,y,t) (\mathcal{I}_m v_0 - v_0)(y) dy}{\int_{\mathbb{R}} \mathcal{K}(x,y,t)( \mathcal{I}_m  v_0)(y) dy} - \frac{2\kappa \int_{\mathbb{R}}   \mathcal{K}(x,y,t) \partial_y(\mathcal{I}_m v_0 - v_0)(y) dy}{\int_{\mathbb{R}} \mathcal{K}(x,y,t)( \mathcal{I}_m  v_0)(y) dy}\\
		&=:&I_1+ I_2.
	\end{eqnarray*}
	Then by the fact  that 
	$u_0\in W^{1,\infty}$ and properties of 
	of the heat kernel, we have 	$u \in W^{1,\infty}$ and also \begin{eqnarray*}
	   \norm{I_1}_{L^2(-\pi,\pi)} \leq C \norm{\mathcal{I}_m v_0-v_0}_{L^2(-\pi,\pi)},\quad 	 \norm{I_2}_{L^2(-\pi,\pi)} \leq C \norm{\mathcal{I}_m v_0-v_0}_{H^1(-\pi,\pi)}.
	\end{eqnarray*}
	We observe that 
	\begin{eqnarray*}
		\Big| \partial_x v_0(s)\Big| &=& \Big| v_0(s)   \frac{1}{2\kappa} u_0(s)\Big|\leq \frac{M_0}{2\kappa} \Big| v_0(s)  \Big| ,\\
		\Big| \partial_{xx} v_0(s)\Big| &=& \Big|\Big(\frac{u_0^2(s)}{4\kappa^2}-\frac{\partial_x u_0(s)}{2\kappa}\Big)   v_0(s)\Big| \leq \Big(\frac{M_0^2}{4\kappa^2} + \frac{M_1}{2\kappa}\Big) \Big| v_0(s)\Big|.
	\end{eqnarray*}
	By the estimates of the  Fourier  interpolation error    
	\begin{eqnarray*}
		\norm{\mathcal{I}_m v_0- v_0}_{H^1(-\pi,\pi)}\leq
		C h\norm{\partial_x^2 v_0}_{L^2(-\pi,\pi)}\leq C(M_0,M_1,\kappa) h,
	\end{eqnarray*}
	we then have 
	$\norm{I_1}_{L^2(-\pi,\pi)}\leq C h$ and $\norm{I_2}_{L^2(-\pi,\pi)}\leq Ch$ and whence for some small $h$ (large $m$)
	\begin{eqnarray*}
		\| \mathcal{G}(u_0)(\cdot,t) - \cG_m  (\mathbf{u}_{0,m})(\cdot,t) \|_{L^2(-\pi,\pi)} \leq C h.
	\end{eqnarray*}
	Here $C>0$ depends on $t$,  $M_0$, $M_1$, and $\kappa$
	and $C$ may be very large when $\kappa$ is small.  
\end{proof}

\section{Dataset size}

\begin{table}[htbp]
    \centering
    \begin{tabular}{l|cc}
    \toprule
    & No. of training data & No. of testing data \\
    \midrule
    \cref{sec:burgers} & 1000 & 200 \\
    \cref{sec:darcy_rectangular} PWC & 1000 & 200  \\
    \cref{sec:darcy_rectangular} Cont. &  1000 & 200  \\
    \cref{sec:darcy_pentagram} & 1900  & 100\\
    \cref{sec:darcy_triangular} & 1900 & 100  \\
    \cref{sec:darcy_notch} & 1900 & 100 \\
    \cref{sec:electroconvection} Electroconvection & 15  & 2  \\
    \cref{sec:advection} Advection I & 1000 & 1000 \\
    \cref{sec:advection} Advection II/III  & 1000 & 200  \\
    \cref{sec:instability_wave} & 40800 & 10000 \\
    \cref{sec:euler} Euler &  100 &  50 \\
    \cref{sec:flapping_airfoil} Airfoil & 28  & 2  \\
    \cref{sec:ns} NS & 1000 & 200   \\
    \cref{sec:cavity} Cavity (Steady) & 100  & 10 \\
    \cref{sec:cavity} Cavity (Unsteady I) &  90 & 10 \\
    \cref{sec:cavity} Cavity (Unsteady II) & 10  & 10\\
    \bottomrule
    \end{tabular}
    \caption{\textbf{Dataset size for each problem, unless otherwise stated.}}
    \label{tab:data_size}
\end{table}

\section{Architectures of DeepONets}

We list the architectures of DeepONet and POD-DeepONet for each example in Tables~\ref{tab:DeepONet_architecture} and \ref{tab:POD_DeepONet_architecture}.

\begin{table}[htbp]
\centering
\begin{tabular}{l|ccc}
\toprule
    &   Branch net  &  Trunk net & Activation \\
    &     &   & function \\
\midrule
\cref{sec:burgers} Burgers & Depth 4 \& Width 128 & Depth 4 \& Width 128 & $\tanh$ \\
\cref{sec:darcy_rectangular} PWC & CNN & Depth 5 \& Width 128 & ReLU \\
\cref{sec:darcy_rectangular} Cont. & CNN & [128, 128, 100] & tanh\\
\cref{sec:darcy_pentagram} &  [128, 128]  &  [128, 128, 128, 128] & tanh \\
\cref{sec:darcy_triangular} &  [128, 128]  &  [128, 128, 128] & ReLU \\
\cref{sec:darcy_notch} & [128, 128]   &  [128, 128, 128]  &  ReLU\\
\cref{sec:electroconvection} Electroconvection &  [256, 256, 128]   &  [128, 128, 128] & ReLU \\
\cref{sec:advection} Advection I &  Depth 2 \& Width 256  &  Depth 4 \& Width 256  & ReLU \\
\cref{sec:advection} Advection II/III & Depth 2 \& Width 512 & Depth 4 \& Width 512 & ReLU \\
\cref{sec:instability_wave} & Depth 6 \& Width 200 & Depth 7 \& Width 200 & ELU \\
\cref{sec:euler} Euler & Depth 2 \& Width 256  &  Depth 4 \& Width 256 & ReLU \\
\cref{sec:flapping_airfoil} Airfoil & Depth 2 \& Width 200  &  Depth 4 \& Width 200  &  ReLU \\
\cref{sec:ns} NS &  CNN  &   [128, 128, 64] & tanh \\
\cref{sec:cavity} Cavity & CNN   & [128, 128, 128, 100]  & tanh \\
\bottomrule
\end{tabular}
\caption{\textbf{DeepONet architecture for each problem, unless otherwise stated.}}
\label{tab:DeepONet_architecture}
\end{table}

\begin{table}[htbp]
    \centering
    \begin{tabular}{l|cc}
    \toprule
         & Branch net & No. of POD modes \\
    \midrule
    \cref{sec:burgers} Burgers & Depth 3 \& Width 128 & 32 \\
    \cref{sec:darcy_rectangular} PWC & CNN & 115 \\
    \cref{sec:darcy_rectangular} Cont. & CNN & 10 \\
    \cref{sec:darcy_pentagram} & [64, 64] & 8 \\
    \cref{sec:darcy_triangular} &  Depth 3 \& Width 128  & 32 \\
    \cref{sec:darcy_notch} & CNN  &  20 \\
    \cref{sec:electroconvection} Electroconvection & Depth 3 \& Width 256  & 12  \\
    \cref{sec:advection} I & Depth 2 \& Width 256 & 38 \\
    \cref{sec:advection} II & Depth 2 \& Width 512 & 38 \\
    \cref{sec:advection} III & Depth 2 \& Width 512 & 32 \\
    \cref{sec:instability_wave} & Depth 6 \& Width 256 & 9 \\
    \cref{sec:euler} Euler &  Depth 2 \& Width 256 &  16 \\
    \cref{sec:ns} NS & CNN & 29 \\
    \cref{sec:cavity} & CNN & 6 \\
    \bottomrule
    \end{tabular}
    \caption{\textbf{POD-DeepONet architecture for each problem, unless otherwise stated.} The activation functions are the same as those in Table~\ref{tab:DeepONet_architecture}.}
    \label{tab:POD_DeepONet_architecture}
\end{table}

\section{Data generation for Darcy flows}
\label{sec:darcy_data}

We employ the {\emph{PDEtoolbox}} in Matlab to solve Eq. \eqref{eq:darcy} for the Darcy flows in different geometries of Sec. \ref{sec:darcy}. 
\begin{itemize}
\item {\bf Rectangular domain:} The Dirichlet boundary conditions are imposed on the inlet and the outlet, i.e., $h(x=0, y) = 1$, and $h(x=1, y) = 0$, and the Neumann boundary conditions are employed on the upper and bottom walls.  We employ 1,893 unstructured meshes in our simulations, and then interpolate the solution to a $20 \times 20$ uniform grid to make it consistent with the generated permeability field. In addition, we obtain 1,200 solutions with different permeability fields, and use 1,000 of them for training and 200 for testing.

\item {\bf Pentagram with a hole:} The Dirichlet boundary conditions are imposed on all boundaries, which are generated from the Gaussian processes in Eq. \eqref{eq:Gaussian_boundary}. We employ 1,938 unstructured meshes in our simulations, and obtain solutions for 2,000 different boundary conditions. We then use 1,900 of them for training the operator networks, and utilize the remaining for testing.  

\item {\bf Triangular domain:} We utilize the same Gaussian processes in Eq. \eqref{eq:Gaussian_boundary} to generate the Dirichlet boundary conditions for all boundaries here, and then employ 861 unstructured meshes in our simulations. Similarly, we generate 2,000 solutions with different boundary conditions, and use 1,900 of them for training and 100 for testing. 

\item {\bf Triangular domain with a notch:} We employ 1,084 unstructured meshes in our simulations to generate 2,000 solutions with different boundary conditions. From the generated datasets, 1,900 samples are used for training, while the remaining 100 are employed as testing samples. 
\end{itemize}

\begin{figure}[htbp]
    \centering
    \subfigure[]{\label{fig:darcy_data1}
    \includegraphics[width=0.45\textwidth]{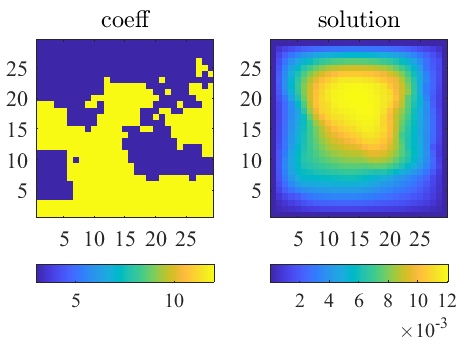}}
    \subfigure[]{\label{fig:darcy_data2}
    \includegraphics[width=0.45\textwidth]{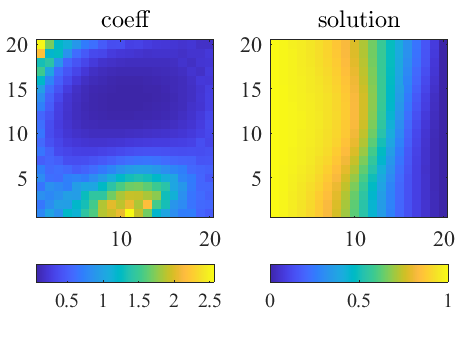}}
    \caption{\textbf{Two datasets of Darcy flow.} (a) Data from \cite{li2020fourier}. (b) Newly-generated data. 
    }
    \label{fig:darcy_testing}
\end{figure}

\begin{figure}
\centering
\subfigure[]{
\includegraphics[width=0.23\textwidth]{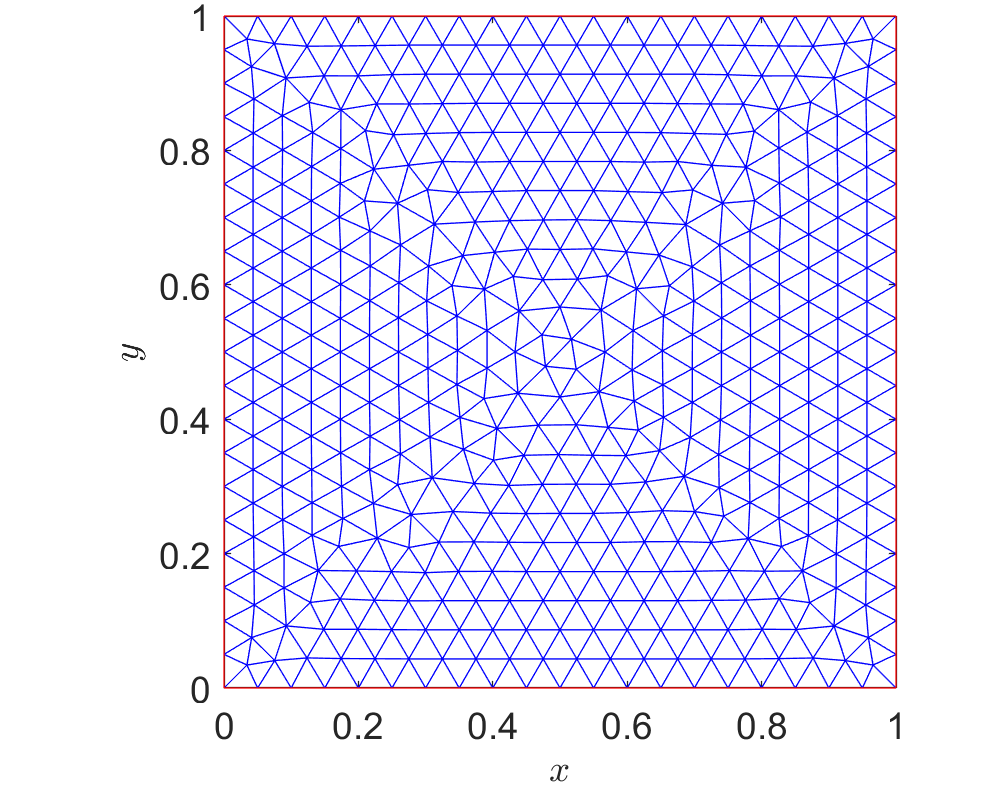}}
\subfigure[]{
\includegraphics[width=0.23\textwidth]{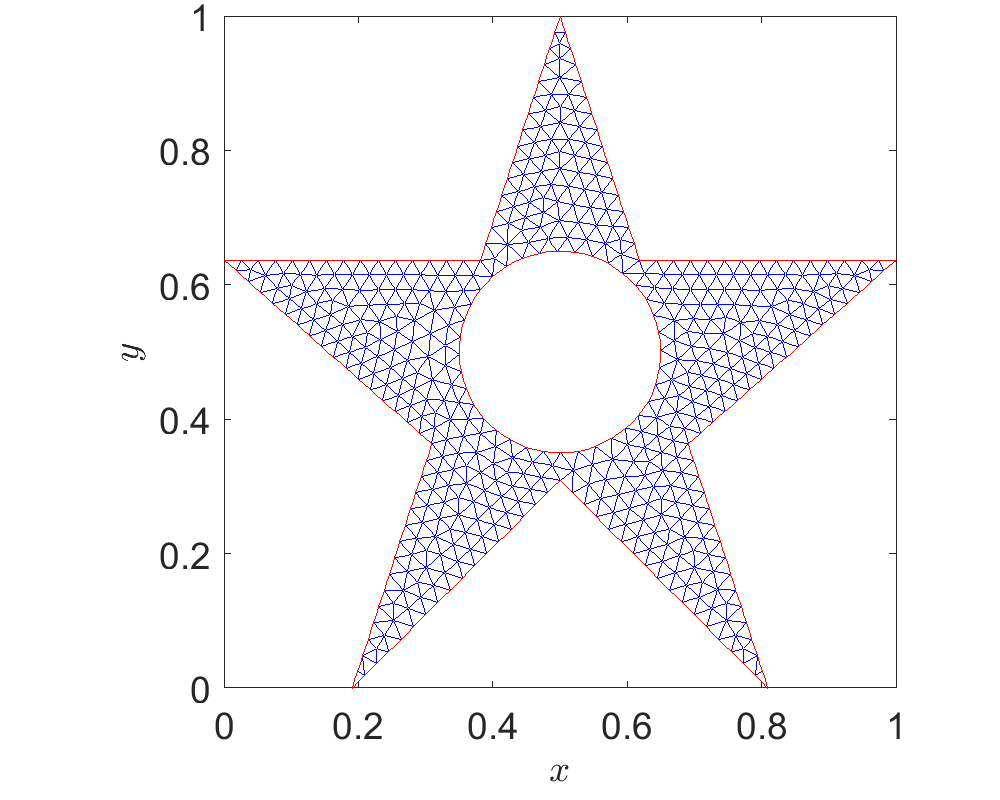}}
\subfigure[]{
\includegraphics[width=0.23\textwidth]{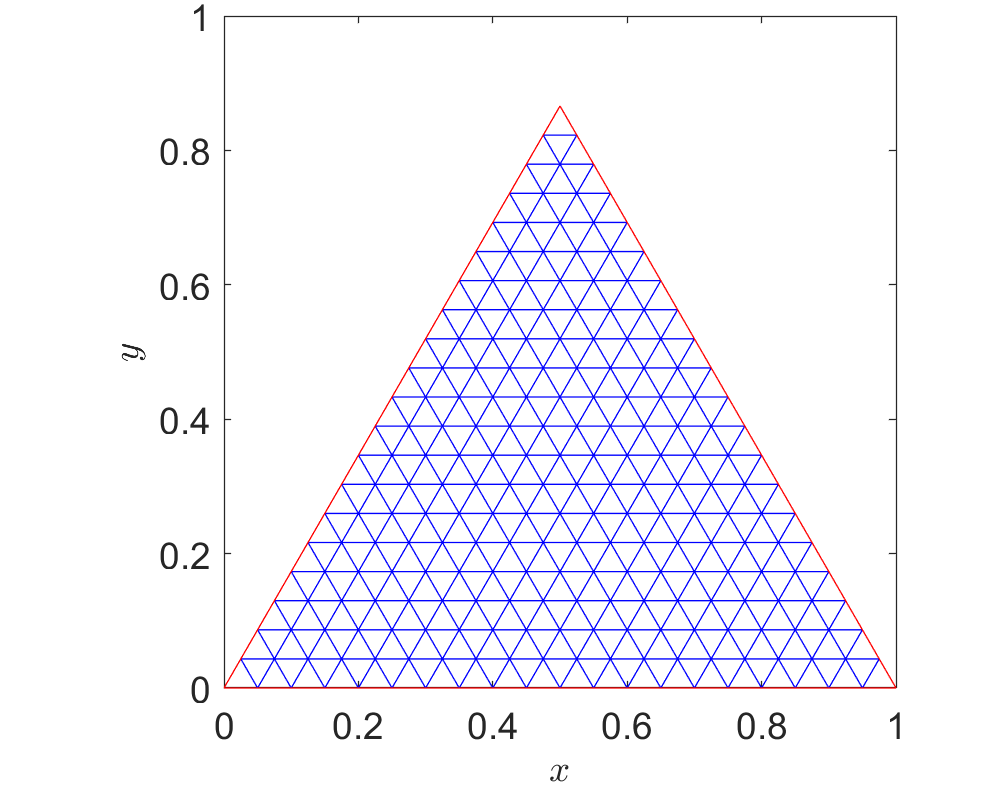}}
\subfigure[]{
\includegraphics[width=0.23\textwidth]{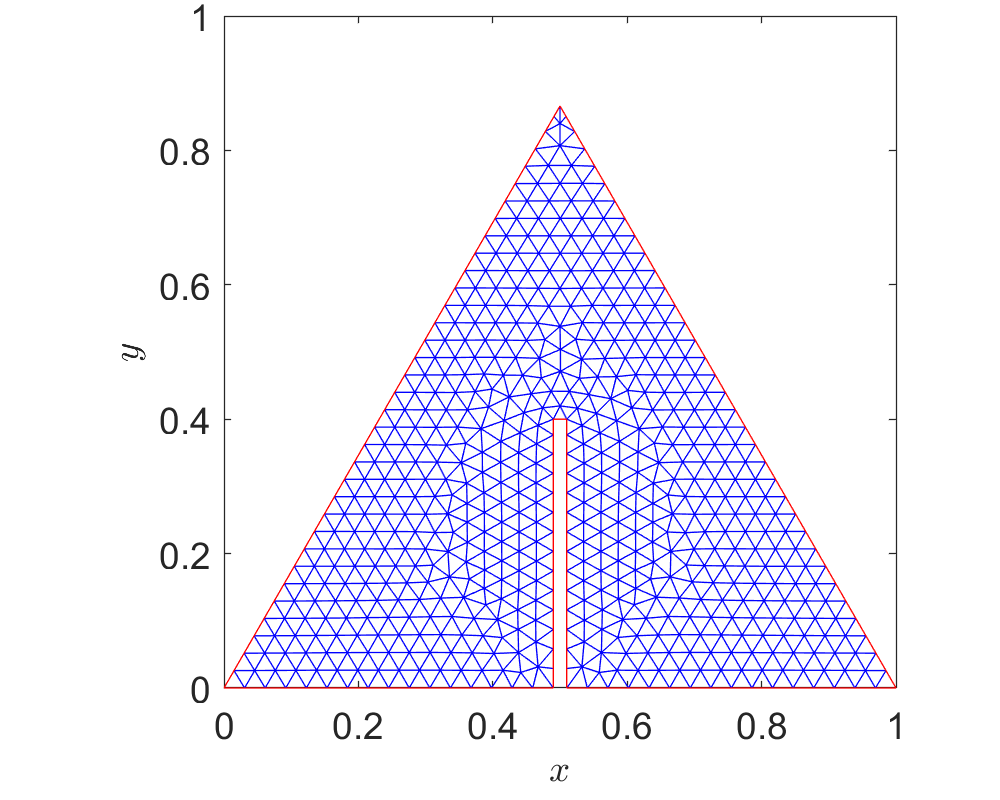}}
\caption{
\textbf{Darcy flows: Unstructured meshes for solving Eq. \eqref{eq:darcy} in different geometries.}
}
\label{fig:darcy_mesh}
\end{figure}

\section{Data generation for regularized cavity flow}
\label{sec:lbm}
For both the steady and unsteady cases, we employ the multi-relaxation-time lattice Boltzmann equation model (MRT-LBE) \cite{meng2015multiple} with a $N \times N = 256 \times 256$ uniform grid in our simulations. In addition, the non-equilibrium extrapolation \cite{zhao2002non} is employed for all boundary conditions. The time step is set as $dt = dx = L/N$.

For the steady case (Case A), we terminate the iteration as the following criterion is met:
\begin{equation*}
    E = \frac{\sum |\bm{u}_{T + 1000} - \bm{u}_T|}{\sum \bm{u}_T} \le 10^{-6},
\end{equation*}
where $T$ denotes the number of iterations.
For the unsteady case (Case B), we set the maximum iteration number as $1,500,000$, and use the velocity fields in the last 10,000 iterations as the training/testing dataset in the DeepONet/FNO.